\documentclass[numberedappendix,apj,chicago]{emulateapj}
\newcommand{\tto}[2]{$#1 \times 10^{#2}$}

\shorttitle{Multidimensional Chemical Modeling. III. Abundance and excitation of diatomic hydrides}
\shortauthors{S. Bruderer et al.}

\journalinfo{Draft version July 19, 2010 (Accepted by ApJ)}
\submitted{Draft version July 19, 2010 (Accepted by ApJ)}

\begin{document} 
\title{Multidimensional chemical modeling of young stellar objects.\\ III. The influence of geometry on\\ the abundance and excitation of diatomic hydrides}

\author{S. Bruderer\altaffilmark{1}}\email{simonbr@astro.phys.ethz.ch}
\author{A.O. Benz\altaffilmark{1}}
\author{P. St\"auber\altaffilmark{1}}
\author{S. D. Doty\altaffilmark{2}}
\altaffiltext{1}{Institute of Astronomy, ETH Zurich, CH-8093 Zurich, Switzerland}
\altaffiltext{2}{Department of Physics and Astronomy, Denison University, Granville, OH  43023, USA}

\begin{abstract}
The Herschel Space Observatory opens the sky for observations in the far infrared at high spectral and spatial resolution. A particular class of molecules will be directly observable; light diatomic hydrides and their ions (CH, OH, SH, NH, CH$^+$, OH$^+$, SH$^+$, NH$^+$). These simple constituents are important both for the chemical evolution of the region and as tracers of high-energy radiation. If outflows of a forming star erode cavities in the envelope, protostellar far UV (FUV; $6 < E_\gamma < 13.6$ eV) radiation may escape through such low-density regions. Depending on the shape of the cavity, the FUV radiation then irradiates the quiescent envelope in the walls along the outflow. The chemical composition in these outflow walls is altered by photoreactions and heating via FUV photons in a manner similar to photo dominated regions (PDRs).

In this work, we study the effect of cavity shapes, outflow density, and of a disk with the two-dimensional chemical model of a high-mass young stellar object introduced in the second paper in this series. The model has been extended with a self-consistent calculation of the dust temperature and a multi-zone escape probability method for the calculation of the molecular excitation and the prediction of line fluxes.

We find that the shape of the cavity is particularly important in the innermost part of the envelope, where the dust temperatures are high enough ($\gtrsim 100$ K) for water ice to evaporate. If the cavity shape allows FUV radiation to penetrate this hot-core region, the abundance of FUV destroyed species (e.g. water) is decreased. On larger scales, the shape of the cavity is less important for the chemistry in the outflow wall. In particular, diatomic hydrides and their ions CH$^+$, OH$^+$ and NH$^+$ are enhanced by many orders of magnitude in the outflow walls due to the combination of high gas temperatures and rapid photodissociation of more saturated species. The enhancement of these diatomic hydrides is sufficient for a detection using the HIFI and PACS instruments onboard Herschel. The effect of X-ray ionization on the chemistry is found to be small, due to the much larger luminosity in FUV bands compared to X-rays.
\end{abstract}
	

%
% Sec: Introduction
%
\section{Introduction} \label{sec:intro}

Observations of young stellar objects (YSOs) envelopes reveal a rich variety of morphologies (e.g. \citealt{Arce06}, \citealt{Jorgensen07a}). Many of the observed features are associated with outflows: For example, strong high-$J$ CO emission along a cavity, etched out by the outflow, is found in the class I object HH 46 (\citealt{vanKempen09a}). They explain the CO narrow emission with a warm and FUV heated region, with FUV photons from either the bow shock, internal jet working surfaces or the accretion disk boundary layer. In this scenario, FUV radiation may travel freely in the low-density cavity without being absorbed or scattered. Similarly, \citet{Bruderer09c} have observed molecular tracers for warm, dense and FUV irradiated gas along the outflow of the high-mass young stellar object AFGL 2591. Since the line shape does not show clear signs of shocks, they suggest the FUV radiation to be of protostellar origin. Indeed, the hot photosphere of a luminous young O or B star emits mostly at FUV wavelengths and a cavity that allows protostellar radiation to escape has also been observed in that source (\citealt{Preibisch03}).

This scenario of irradiated outflow walls along a cavity, swept out by an outflow, has also been used to explain other observations. For example, \citet{Jorgensen04b} found emission of FUV enhanced CN along an outflow of a low-mass class 0 source. Van Kempen et al. (\citeyear{vanKempen09b}) explain the water maser emission at 183 GHz in Ser SMM 1 with the increased water abundance in the warm, FUV or shock heated outflow walls. Outflow walls can also be sites of anomalous molecular excitation. For example \citet{Hogerheijde98,Hogerheijde99} observe strong emission of HCN and HCO$^+$ along the outflow of L 1527 and Ser SMM 1. These molecules have a large dipole moment and are efficiently excited by collisions with electrons, mixed from the ionized outflow into the outflow wall. \citet{Rawlings04} explain the strong HCO$^+$ emission by shock liberation of molecules from ice mantles followed by photoprocessing. \citet{Spaans95} model the strong $^{13}$CO $J=6 \rightarrow 5$ emission in narrow lines observed toward several low-mass sources (e.g. TMR-1) by FUV heated outflow walls. \citet{Bruderer09b} (second paper of this series, hereafter paper II) explain the observed high abundance of the FUV enhanced molecule CO$^+$ in AFGL 2591 by a concave cavity that allows direct irradiation onto the outflow walls.

Diatomic hydrides and their ions (XH, XH$^+$) are corner-stones of the chemical network. For example water at high temperature ($\gtrsim 250$ K) is mainly formed by the reaction of OH + H$_2$. Due to their chemical properties, discussed in this work, they trace dense, warm and FUV or X-ray irradiated gas like in outflow walls. Such tracers are valuable to constrain the FUV or X-ray radiation of embedded sources. This is of particular interest for the innermost part of low-mass YSOs with protoplanetary disks being ionized and heated by FUV radiation and X-rays (e.g. \citealt{Bergin09}). Here, we discuss the use of light diatomic hydrides with the common elements (X=O, C, N, S) as tracers of protstellar high-energy radiation. The chemistry of other diatomic hydrides with heavier (e.g. X=Fe) and/or less abundant elements (e.g. X=F) is not discussed here, as their chemistry is poorly known and their use as tracers currently limited. However, with the ongoing search for new diatomic hydrides using the new Herschel Space Observatory (e.g.\citealt{Cernicharo10}), this sitation may change. For example \citet{Neufeld10} report the detection of HF using the HIFI instrument onboard Herschel. Due to the dissociation energy of HF being larger than of H$_2$, HF can be formed in an exothermic reaction of F with H$_2$ and HF may become the dominant reservoir of fluorine (\citealt{Neufeld09}). The silicon bearing SiH and SiH$^+$ may be interesting in the context of shocked regions where ``sputtering'' from dust grains (e.g. \citealt{Flower96}) can increase the gas phase abundance of silicon.

The Herschel Space Observatory and Band 10 receivers of the upcoming Atacama Large Millimeter and submillimeter Array (ALMA) allow the study of important FUV and X-ray tracers, like diatomic hydrides, at high sensitivity and with good spatial and spectral resolution for the first time. This raises the questions: What is the expected abundance and line flux of FUV enhanced species in the scenario of directly irradiated outflow walls? From which region of the envelope does the molecular radiation emerge and how are the molecules excited? How dependent are observable quantities (line fluxes or line ratios) on characteristics that cannot be constrained directly from observations (e.g. geometry in the innermost part of the envelope and chemical age)? To attack such questions, detailed numerical models of the chemistry of YSOs envelopes can be used.

Models of the chemistry of YSO envelopes have so far mostly assumed a slab or spherical symmetry and a static physical structure (e.g. \citealt{Viti99a}, \citealt{Doty02}). In situations with chemical timescales longer than dynamical timescales, the chemistry needs to be coupled with the dynamical evolution (e.g. \citealt{Doty06}). A particular example for this situation is the slow sulfur chemistry (\citealt{Wakelam04a}). Other models focus on the effect of protostellar FUV and X-ray radiation on the chemistry of a spherical envelope (\citealt{Staeuber04,Staeuber05}). Due to the high extinction of FUV radiation by dust, they find the chemistry to be modified only in the innermost few hundred AU. X-rays on the other hand, have a much lower absorption cross section. They may thus penetrate further into the envelope and dominate the ionization in the entire hot core. To study the effects of an outflow cavity that allows FUV radiation to escape and irradiate a larger part of the envelope, two-dimensional axisymmetric models are necessary.

Non-spherical chemical models of YSO envelopes are presented by \citet{Brinch2008a}, \citet{Visser09a} and in paper II. These models assume axisymmetry. The first two are applied on low-mass sources and include the effects of infall to study the chemical evolution from the cloud to the disk. The model in paper II is applied on a high-mass star forming region. It calculates the gas temperature self-consistently with the chemical abundances similar to models of photo dominated regions (PDRs; e.g. \citealt{Kaufman99}, \citealt{Meijerink05a}), but it does not account for the dynamical evolution. The model presented in paper II considers FUV ($6 < E_\gamma < 13.6$ eV) and X-ray ($E_\gamma > 100$ eV) induced chemistry.

In this work, we further develop the model introduced in paper II. It is based on the grid of chemical models introduced in the first paper of this series of papers (\citealt{Bruderer09a}, hereafter paper I). We extend the model with two important tools: (i) with a dust radiative transfer method for a self-consistent calculation of the dust temperature and (ii) a multi-zone escape probability method to study the molecular excitation and radiation. We apply the model to AFGL 2591 to answer the questions raised above. In particular we predict line fluxes of two selected diatomic hydrides (SH and CH$^+$), having typical properties of diatomic hydrides with high temperatures needed for formation and high critical densities. These two molecules will be observed in AFGL 2591 by the Water In Star-forming regions with Herschel (WISH\footnote{www.strw.leidenuniv.nl/WISH}; van Dishoeck et al. 2010, in prep.) guaranteed time key program.

This paper is organized as follows. We first describe the details of the model in section \ref{sec:model}. The physical structure of different models is discussed in section \ref{sec:physstruc}. The chemical abundances obtained from these physical structures are presented in section \ref{sec:chemabu}, and compared to the results of one-dimensional models. The molecular excitation and line radiation is then studied in section \ref{sec:molline}, and the scientific prospectives of hydride observations with Herschel are discussed in \ref{sec:herschobs}. Limitations of the model used in this work are given in Section \ref{sec:limit}.

%
% Sec: Method
%
\section{Model} \label{sec:model}

In this section, we briefly describe the calculation flow of the models used here. Details on the new parts of the model, the dust radiative transfer and the molecular excitation calculation, are given in Appendicies \ref{sec:dustrt} and \ref{sec:molexcit}, respectively.

The modeling process starts with assuming a density structure and the spectrum of the central source. A dust radiative transfer calculation then solves for the dust temperature. The local FUV and X-ray radiation is calculated at every point of the model as described in Appendix \ref{sec:dustrt}. In the next step, the gas temperature is obtained by equating the heating and cooling rates (paper II). Since these rates depend on the chemical abundances (e.g. of the atomic coolants C$^+$ and O) the calculation of the gas temperature has to be performed simultaneously with the chemistry. This time-consuming step is facilitated by the grid of chemical models introduced in paper I. The grid consists of a database of precalculated abundances depending on physical parameters (density, temperature, FUV flux, X-ray flux, cosmic-ray ionization rate and chemical age; see paper I for a definition of these parameters) from which abundances can be interpolated quickly. The  interpolation approach is also used to obtain the abundances of all other species (e.g. diatomic hydrides). The molecular excitation is calculated using the escape probability approach described in Appendix \ref{sec:molexcit} and yields the modeled line radiation, thus the observable quantities.

\subsection{A grid of density structures/cavity shapes} \label{sec:griddensity}

A spherical model of AFGL 2591 has been constructed by \citet{vdTak99,vdTak00a} based on maps of the dust continuum, the spectral energy distribution (SED) and molecular line emission. In their fitting, they find good agreement on larger scales. However, continuum emission at shorter wavelengths (warm dust) requires a shallower density profile in the innermost $10''-20''$. They conclude that an inner region with roughly constant temperature may be required to reproduce all observations. \citet{Jorgensen05c} encountered similar problems of an excess emission in short wavelengths when fitting the continuum emission of IRAS 16293-2422. They concluded that a large spherical cavity can explain this excess. However, \citet{Crimier10} can explain both SED and maps by a higher bolometric luminosity. In AFGL 2591, new high resolution observations at 24.5 $\mu$m by \citet{deWit09} show extended emission at scales of up to $10''$ in the direction of the outflow. This emission stems from the outflow walls (\citealt{Bruderer09c}), supporting the scenario of a cavity.

\begin{figure}[tbh]
  \centering
  \includegraphics[width=0.8\hsize]{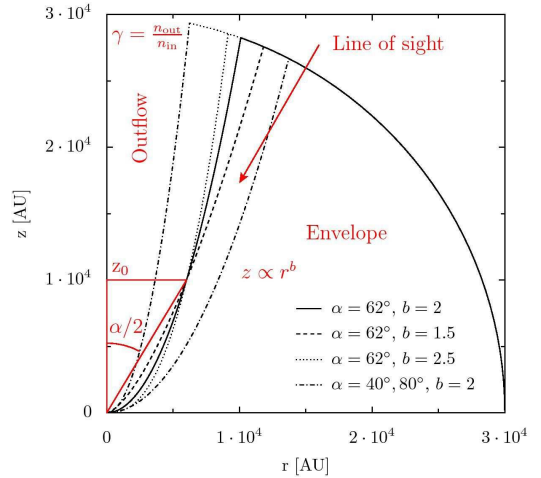}
  \caption{Separation between envelope and outflow using the power-law description (Equation \ref{eq:seppow}). The line of sight suggested by \citet{vdTak99} is given by an arrow.\label{fig:separation}}   
\end{figure}

For this work, we adopt the density structure of the spherical model by \citet{vdTak99}. This choice allows to study the differences of the two-dimensional chemical model including an outflow cavity with the spherically symmetric models by \citet{Doty02} and \citet{Staeuber04,Staeuber05}. The cavity shape is assumed to follow a  power-law with index $b$,
\begin{equation} \label{eq:seppow}
z = \frac{1}{z_0^{b-1} \tan(\alpha/2)^b} r^b \ ,
\end{equation}
where $z$ and $r$ are coordinates along the outflow axis and perpendicular to the outflow, respectively. The full opening angle $\alpha$ of the outflow is given at a distance $z_0$. As in paper II, we use $z_0=10000$ AU, approximately the scale on which \citealt{Preibisch03} have probed the opening angle of the outflow cavity. Cavity shapes for different choices of $\alpha$ and $b$ are given in Figure \ref{fig:separation}. The solid line indicates the model used in paper II. Figure \ref{fig:separation} also shows the line of sight toward us crossing the outflow wall at a shallow angle, as suggested by \citet{vdTak99}. A further parameter of the model is the density in the cavity, which we assume to be reduced by a constant factor $\gamma=n_{\rm out}/n_{\rm in}$ compared to the spherical density model at the same radius. This is motivated by the pressure equilibirum along the outflow wall (paper II). Note that for $\gamma=2.5 \times 10^{-3}$, used in most of the models, the results are independent of that assumption as the opacity in FUV wavelengths along the outflow is low.

A grid of different models is discussed in the following sections. Starting from the model used in paper II (referred as model \textit{Standard} in the following) we explore the parameter range by varying the outflow opening angle $\alpha$, the power-law index $b$, cavity density $\gamma$ and the bolometric luminosity of the protostar $L_{\rm bol}$ (Table \ref{tab:modelsafgl}). As in paper II, we assume the central source to emit a black body spectrum with a temperature of \tto{3}{4} K following \citet{vdTak99} (see Section \ref{sec:limit}). Thus, the FUV field has photons that can dissociate CO and H$_2$.

\begin{table}[tbh]
\caption{Parameters of the adopted models.}
\label{tab:modelsafgl}
\centering
\begin{tabular}{ll|llll}
\hline
& Model       & $\alpha$ & $b$ & $\gamma=n_{\rm out}/n_{\rm in}$ & $L_{\rm bol}$ \\
&            & [${}^{\circ}$] &                           &     & [$L_\odot$]   \\
\hline
&     \textit{Standard}       & 62       & 2 &    2.5(-3)                         & 2(4) \\
&     \textit{Disk$^a$}       & 62       & 2 &    2.5(-3)                         & 2(4) \\
\multicolumn{2}{l|}{Opening angle} &\\
&     \textit{$\alpha$20}     & 20       & 2 &    2.5(-3)                         & 2(4) \\
&     \textit{$\alpha$40}     & 40       & 2 &    2.5(-3)                         & 2(4) \\
&     \textit{$\alpha$80}     & 80       & 2 &    2.5(-3)                         & 2(4) \\
&     \textit{$\alpha$100}    & 100      & 2 &    2.5(-3)                         & 2(4) \\
\multicolumn{2}{l|}{Cavity density} & \\
&     \textit{$\gamma$0}      & 62       & 2 &    0                               & 2(4) \\
&     \textit{$\gamma$0.01}   & 62       & 2 &    1(-2)                           & 2(4) \\
&     \textit{$\gamma$0.1}    & 62       & 2 &    0.1                             & 2(4) \\
\multicolumn{2}{l|}{Cavity shape} &\\
&     \textit{$b$1.5}         & 62       & 1.5 &  2.5(-3)                         & 2(4) \\
&     \textit{$b$2.5}         & 62       & 2.5 &  2.5(-3)                         & 2(4) \\
\multicolumn{2}{l|}{Source luminosity}& \\
&     \textit{$L$2e3}         & 62       & 2 &    2.5(-3)                         & 2(3) \\
&     \textit{$L$1e4}         & 62       & 2 &    2.5(-3)                         & 1(4) \\
&     \textit{$L$4e4}         & 62       & 2 &    2.5(-3)                         & 4(4) \\
\hline
\end{tabular}
\begin{flushleft}
\footnotesize{
Note: $a(b)=a \times 10^b$\\
$^a$Parameters like in the model \textit{Standard}, but with a disk of mass 0.8 $M_\odot$ in the innermost 400 AU.}
\end{flushleft}
\end{table}

A compact structure with a systematic velocity gradient has been detected in AFGL 2591 by \citet{vdTak06} using interferometric observations. They propose a circumstellar disk to be the source of this radiation and constrain a mass of $\approx0.8$ $M_\odot$ within a radius of 400 AU to the protostar. While a detailed modeling of the disk is beyond the scope of this study, model \textit{Disk} implements a toy-model of a disk embedded in the envelope of the model \textit{Standard}. The toy-model consists of a massless (in relation to the protostar) Keplerian disk (e.g. \citealt{Fischer96}) with a surface density $\Sigma \propto r^{-0.75}$, a scale height $r/z=0.15$ and a mass of $0.8$ $M_\odot$ within a radius of 400 AU to the protostar. The density distribution of the disk and the inner envelope is given in Figure \ref{fig:disk}.

\section{Physical structure} \label{sec:physstruc}

Different physical conditions such as gas and dust temperature, density, FUV and X-ray irradiation govern the chemical composition and molecular excitation. As a first step to quantify these physical properties of the different envelopes models, we study the amount of material with distinct conditions (e.g. a gas temperature above 100 K). Certain conditions like warm temperatures, a high density and a FUV field stronger than 1 ISRF (interstellar radiation field) can be necessary for the formation and excitation of a molecule. Combinations of these regions (e.g. high density and strong FUV field) can be relevant for the formation and excitation of molecules and thus the predicted line strength that can be compared with observations. While the definition of such physical regions is arbitrary, it gives a simple overview on differences between models and thus the influence of the geometry.

In Table \ref{tab:propafgl}, we show the absolute mass (in $M_\odot$) and mass relative to Model \textit{Standard} ($M/M_{\rm standard}$) for the following regions:%
\begin{itemize}
\item \textit{$T_{\rm dust},T_{\rm gas} > 100$ K.} Several molecules only form at high temperature, like for example the diatomic hydrides discussed in this work. We report on the region with dust and gas temperature above 100 K, which is assumed to be the water evaporation temperature (paper I).
\item \textit{FUV.} FUV radiation alters the chemical composition by photoprocesses (ionization and dissociation) and also through heating of the gas. The FUV region is defined by the FUV flux larger than 1 ISRF.  We recall, however, that the structure of the outflow walls is more complicated and some species only exist in different layers, for example the C$^+$/C/CO layer structure also seen in PDRs. 
\item \textit{X-rays.} X-rays affect the chemistry mostly by secondary electrons created by fast photoelectrons ionizing H$_2$ (\citealt{Maloney96,Staeuber05}). Assuming an X-ray luminosity of $10^{32}$ erg s$^{-1}$ in the 0.1 - 100 keV band as suggested by \citet{Staeuber05}, heating by X-rays can be neglected (paper II and \citealt{Staeuber05}). While the influence of X-rays is very similar to that of a high cosmic-ray ionization rate (paper I), there are differences in FUV irradiated regions (e.g. \citealt{Staeuber06} for a discussion of the water abundance in FUV and X-ray irradiated regions). The X-ray region is defined by the H$_2$ ionization rate larger than the ``standard'' cosmic-ray ionization rate of $\zeta_{\rm c.r.} = 5.6 \times 10^{-17}$ s$^{-1}$ by a factor of about five ($\zeta > 3 \times 10^{-16}$ s$^{-1}$). This anticipates the uncertainty in the observational determination of the cosmic-ray ionization rate (\citealt{vdTak00b}, \citealt{Doty02}, \citealt{Indriolo07}).
\item \textit{$T > 100$ K \& No FUV.} An important region of the envelope is the part which has a temperature above 100 K but no FUV irradiation. Hot core species which can be photodissociated may exist only in that region. The region is defined by the temperature $T=\max(T_{\rm dust},T_{\rm gas})$ (see Section \ref{sec:chemabu}) and the FUV region, defined above.
\end{itemize}
The spherical model is also given in Table \ref{tab:propafgl}. For comparison with earlier modeling, we follow \citet{Staeuber04,Staeuber05} and assume a visual extinction of $\tau_V=0$ and $G_0=10$ ISRF at the inner edge of the modeled region at 200 AU and use the gas and dust temperature by \citet{Doty02}.

For the model \textit{Spherical}, a plot of the radial dependence of density and temperature is given in Figure 3 of paper I. For this model, the radial size of the regions discussed here are: 1900 AU (\textit{$T_{\rm dust} > 100$ K}), 1100 AU (\textit{$T_{\rm gas} > 100$ K} and \textit{$T > 100$ K (No FUV)}), 224 AU (\textit{FUV}), 2300 AU (\textit{X-rays} and \textit{only X-rays}). This shows the small size of the FUV irradiated and heated region in a spherical model compared to the models including and outflow wall. The extent of the different regions of model \textit{Standard} is indicated in Figure \ref{fig:xrayfuv} and the inset of Figure \ref{fig:geomdiff}.

The X-ray luminosity of AFGL 2591 cannot be measured directly do to the high column density towards the protostar, but upper limits of $1.6 \times  10^{31}$ erg s$^{-1}$ are given in \citet{Carkner98}. However, they assume an attenuating column density to the source of only $N_{\rm H} = 10^{22}$ cm$^{-2}$, much smaller than the radial column density of about $N_{\rm H} = 3 \times 10^{23}$ cm$^{-2}$ of the density profile used here. Note that the column density derived from the density profile can be affected by uncertainties e.g. in the adopted dust opacity (\citealt{vdTak99}). For the plasma temperature of $7 \times 10^{7}$ K assumed here (paper II), the X-ray luminosity may be a factor of about 10 higher assuming an attenuating column density of $N_{\rm H} = 3 \times 10^{23}$ cm$^{-2}$ (e.g. Appendix A of paper I). We thus consider the adopted value of the X-ray luminosity of $10^{32}$ erg s$^{-1}$ as an upper limit.

\begin{table*}[tbh]
\tablewidth{0pt}
\footnotesize
\centering
\caption{Mass of different regions in the 2D models and the spherical model. The upper part of the table gives absolute values of the mass (in $M_\odot$), while the lower part of the table gives factorial deviations $^a$ relative to the model \textit{Standard}. Factorial deviations larger than a factor of three are highlighted. The total mass of the envelope and the mass within a radius of 4000 AU and 20000 AU to the protostar are  given by $M_{tot}$, $M_{4000}$ and $M_{20000}$, respectively. ``1'' means that the value is by definition equal to the model \textit{Standard}. The spatial extent of the regions is discussed in Section \ref{sec:physstruc}.\label{tab:propafgl}}
\begin{tabular}{llccc|cccccc}
\tableline
 & Model & $M_{tot}$ & $M_{4000}$ & $M_{20000}$ & $T_{\rm dust} > 100$ K & $T_{\rm gas} > 100$ K & FUV & X-rays & only X-rays$^b$ &  $T > 100$ K (No FUV)$^c$ \\
\tableline
\multicolumn{4}{l}{\textbf{Absolute Values ($M_\odot$)}} & & & & & & & \\
  &  \textit{Spherical}      &     49.4 &    0.94 &    33.5  &    0.20    &    6.8(-2) &    5.7(-4) &       0.29 &       0.29 &    6.7(-2) \\
  &  \textit{Standard}       &     44.3 &    0.45 &    18.7  &   6.7(-2) &       0.71 &        1.9 &       0.70 &       0.32 &    7.9(-3) \\
\tableline
\multicolumn{4}{l}{\textbf{Relative to Model \textit{Standard}}} & \\
  &  \textit{Spherical}      &      1.1 &    2.1  &     1.8  &    \textbf{3.0}   &   \textbf{9.6(-2)} &    \textbf{3.0(-4)} &       0.42 &       0.92 &      \textbf{8.5} \\
  &  \textit{Standard}       &      1   &    1    &     1    &     1   &        1   &        1   &        1   &        1   &        1   \\
  &  \textit{Disk}           &      1.1 &    9.1  &     1.2  &   \textbf{ 15.6 }&        2.3 &        1.2 &        1.3 &       0.82 &       \textbf{83.7} \\
\multicolumn{4}{l}{Opening angle} & \\
  &  \textit{$\alpha$20}     &      1.1 &    1.8  &     1.2  &        2.3 &       0.36 &       \textbf{0.25} &       0.70 &        1.3 &       \textbf{17.0} \\
  &  \textit{$\alpha$40}     &      1.1 &    1.5  &     1.1  &        1.5 &       0.60 &       0.60 &       0.87 &        1.2 &        \textbf{6.7} \\
  &  \textit{$\alpha$80}     &     0.91 &    0.64 &     0.87 &       0.72 &        1.4 &        1.4 &       0.90 &       0.47 &          \textbf{0} \\
  &  \textit{$\alpha$100}    &     0.75 &    0.36 &     0.68 &       0.49 &        1.8 &        1.8 &       0.73 &    \textbf{9.4(-2)} &          \textbf{0} \\
\multicolumn{4}{l}{Cavity density} & \\
  &  \textit{$\gamma$0}      &      1   &    1    &     1    &        1.1 &        1.2 &        1.1 &        1.0   &        1.0   &       0.71 \\
  &  \textit{$\gamma$0.01}   &      1   &    1    &     1    &       0.91 &       0.46 &       0.73 &       0.93 &        1.0   &        1.8 \\
  &  \textit{$\gamma$0.1}    &      1   &    1    &     1    &       0.71 &    \textbf{1.2(-2)} &    \textbf{3.8(-3) }&       0.65 &        1.4 &        \textbf{5.5} \\
\multicolumn{4}{l}{Cavity shape} & \\
  &  \textit{$b$1.5}         &     0.99 &    1.3  &     1.01 &        1.2 &       0.90 &       0.95 &       0.76 &       0.85 &        \textbf{4.3} \\
  &  \textit{$b$2.5}         &      1   &    0.76 &     0.99 &       0.91 &        1.1 &        1.1 &        1.1 &       0.90 &    \textbf{8.9(-2)} \\
\multicolumn{4}{l}{Source luminosity} & \\
  &  \textit{$L$2e3}         &      1   &    1    &     1    &    \textbf{6.9(-2)} &       0.37 &       0.66 &        1   &        1.2 &          \textbf{0} \\
  &  \textit{$L$1e4}         &      1   &    1    &     1    &       0.41 &       0.76 &       0.89 &        1   &        1.0   &    \textbf{8.9(-2)} \\
  &  \textit{$L$4e4}         &      1   &    1    &     1    &        2.3 &        1.3 &        1.1 &        1   &       0.96 &        \textbf{5.0} \\
\tableline
\end{tabular}
\begin{flushleft}
\footnotesize{Note: $a(b)=a \times 10^b$.\\
$^a$Defined by $max(M/M_{\rm standard},M_{\rm standard}/M)>3$.\\
$^b$Only in the X-ray region and not in the FUV region.\\
$^c$Region with $T=\max(T_{\rm gas},T_{\rm dust}) > 100$ K outside the FUV region.}
\end{flushleft}
\end{table*}

\subsection{Total mass}

The total mass of the envelopes ($M_{\rm tot}$) of the various models is very similar with maximum differences of 25 \% compared to model \textit{Standard} (Table \ref{tab:propafgl}). The situation is different if only the inner part of the envelope is considered. We give the mass within a radius of 4000 and 20000 AU to the protostar ($M_{4000}$ and $M_{20000}$) following the Herschel beam at the shortest/longest wavelengths of PACS/HIFI (60-625 $\mu$m) assuming a distance of 1 kpc to the source (\citealt{vdTak99}). Clearly, there is foreground material within the Herschel beam. In this work, however, we concentrate on molecules with lines having high critical densities. They thus trace only the inner regions due to the density gradient. However, dust or line radiation from inner regions may also pump molecules further out (e.g. \citealt{vanKempen08}). A detailed discussion which regions are traced by molecular emission requires the calculation of the excitation as presented in Section \ref{sec:molline}.

Models with a large opening angle (\textit{$\alpha$100}) or the toy-disk have masses  $M_{4000}$ which are 3 times smaller and 10 times larger, respectively, compared to the model \textit{Standard}. More than half of the envelope mass is outside a radius of 20000 AU and differences between individual models are less evident in $M_{20000}$ and not seen in $M_{\rm tot}$.

We conclude that different cavity shapes affect the total mass only in the inner part of the envelope, with the largest changes found for models with a large/small outflow opening angle $\alpha$ or a disk embedded in the envelope.

\subsection{Dust temperature above 100 K} \label{sec:phydust100}

The mass with dust temperature above 100 K approximately scales with $M_{4000}$ for models with the disk and different cavity shape (Models \textit{$\alpha$20-$\alpha$100}, \textit{b1.5}, \textit{b2.5}). This is explained by the dust temperature depending only on the distance to the protostar and not on the density in optically thin, spherically symmetric models. For this case, the radius for a given temperature varies as $L_{\rm bol}^{0.5}$ (e.g. \citealt{Jorgensen06}). For the density structure $\propto r^{-1}$, the mass with $T_{\rm dust} > 100$ K should scale $\propto L_{\rm bol}$. Models \textit{L2e3},  \textit{L1e4} and \textit{L4e4} depend slightly stronger on $L_{\rm bol}$ due to the cavity removing more of the dense material in the inner part compared to larger distances to the protostar. Changing the density in the cavity (Models \textit{$\gamma$0}, \textit{$\gamma$0.01} and \textit{$\gamma$0.1}) affects the mass with $T_{\rm dust} > 100$ K by the attenuation of direct stellar radiation in FUV wavelengths that heats the dust in the outflow walls.

We conclude that the amount of material with dust temperature above 100 K is within a factor of two to the model \textit{Standard} for models with a different cavity shape, but larger for the models including a \textit{Disk} and with a lower bolometric luminosity (\textit{L2e3}).

\subsection{Gas temperature above 100 K} \label{sec:phygas100}

Gas with $T_{\rm gas} > 100$ K is more than an order of magnitude less present in the spherical model and model \textit{$\gamma$0.1} compared to the model \textit{Standard}. This underlines the impact of the empty cavity to extensively enhance warm gas by escaping FUV radiation. Model \textit{$\gamma$0.1} has even less warm gas than the spherical model, due to a more efficient cooling through the low-density outflow region. In all other models, the deviation of the mass with $T_{\rm gas} > 100$ K compared to model \textit{Standard} is smaller than a factor of three. In particular, the models with a different bolometric luminosity (\textit{$L$2e3}, \textit{$L$1e4} and \textit{$L$4e4}) scale less than with $L_{\rm bol}$ due to the PDR surface temperature scaling with less than $G_0$ for densities below about $10^6$ cm$^{-3}$ (e.g. \citealt{Kaufman99}, \citealt{Meijerink07b}).

We conclude that the mass with gas temperatures above 100 K changes most if the geometrical situation is similar to the spherically symmetric model. These are models with a small outflow opening angle or a high density of absorbing material in the outflow.

\subsection{X-ray and FUV irradiation} \label{sec:phyxrayfuv}

\begin{figure}[tbh]
  \centering
  \includegraphics[width=1.0\hsize]{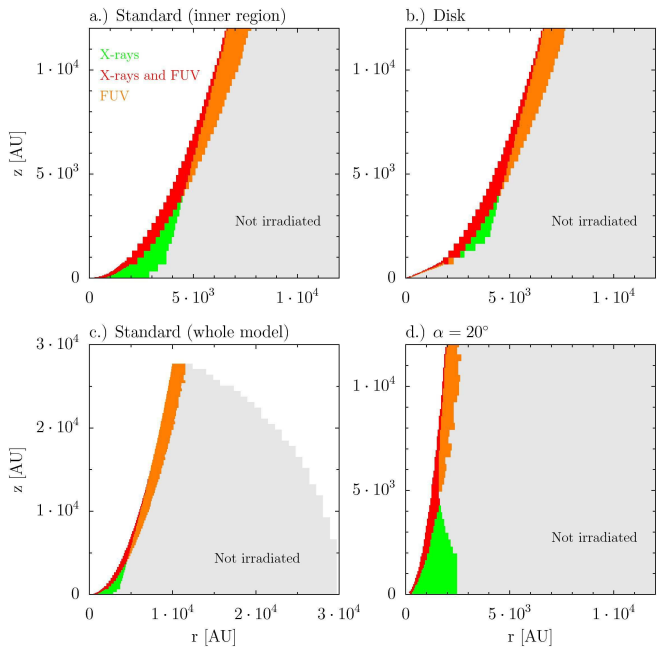}
  \caption{FUV and X-ray irradiated regions in the model \textit{Standard} and the models with a disk and $\alpha=20^\circ$. FUV and X-ray irradiated regions as defined in Section \ref{sec:physstruc} are given in green (only X-rays), orange (only FUV) and red (both X-rays and FUV).}
   \label{fig:xrayfuv}
\end{figure}

Regions with high-energy irradiation (FUV and/or X-rays) are given in Figure \ref{fig:xrayfuv} for models \textit{Standard}, \textit{disk} and \textit{$\alpha$20}. They show an FUV region along the outflow cavity. The X-ray region is less concentrated to the surface and also includes the envelope to distances of up to a few 1000 AU. A region with only X-ray influence follows the FUV irradiated surface layer in the vicinity of the protostar (distance $<5000$ AU). Some surface concentration is also seen in X-rays due to photons with energy between 0.1 and 1 keV which are attenuated by a relatively low column density (of order $10^{22}$ cm$^{-2}$). Harder X-rays with $E_\gamma > 1$ keV require a column density of order $10^{24}$ cm$^{-2}$ to be attenuated (paper I). A disk can provide such a high column density and shield the envelope even for hard X-rays (Figure \ref{fig:xrayfuv}).

The mass of the X-ray/FUV irradiated gas is the same for most models within a factor of two relative to model \textit{Standard}. Exceptions are the FUV irradiated mass in the spherical model and model \textit{$\gamma$0.1} being orders of magnitude smaller. Smaller opening angles (Model \textit{$\alpha$20}) yield a larger incident angle of the FUV radiation and thus a thinner outflow wall. This is reflected in the FUV irradiated mass and also the mass with $T_{\rm gas} > 100$ K. For large opening angles (\textit{$\alpha$100}) the FUV radiation penetrates to the midplane of the flat structure and the region with only X-ray irradiation is much smaller. In the model including a disk, the mass suffering X-ray irradiation is not much different despite the shielding seen in Figure \ref{fig:xrayfuv}. It is caused by the larger density in the innermost part at the surface of the disk.

The largest variations between different models are found for the mass of the region with $T > 100$ K but no FUV irradiation. This region is shown in yellow in the Figures \ref{fig:geomdiff} and \ref{fig:disk} for models \textit{Standard}, \textit{b1.5}, \textit{b2.5}, \textit{$\alpha$20} and \textit{disk}. A thin layer of FUV heated gas is followed by an infrared heated region. If the region close to the protostar is geometrically thin (\textit{b2.5}, \textit{$\alpha$80} and \textit{$\alpha$100}) FUV radiation can penetrate to the midplane and no material with $T > 100$ K but no FUV irradiation exists. Geometrically thin/thick means that the distance in vertical direction between the midplane and the outflow is small/large. In models with a geometrically thick inner region (\textit{b1.5}, \textit{$\alpha$80} and \textit{$\alpha$100}) or more mass in the innermost region (\textit{disk}) the amount of IR heated gas with $T > 100$ can be even larger than in the spherically symmetric model. The strong sensitivity of this mass on the bolometric luminosity is the result of the dust mass with temperature above 100 K depending more on the bolometric luminosity than the gas temperature (Section \ref{sec:phydust100} and \ref{sec:phygas100}).

We conclude that the mass with X-ray irradiation is not considerably affected by the geometry, while the mass with FUV irradiation shows differences for those models that are similar to the spherical situation (small opening angle or high density of absorbing material in the outflow). The region with temperature above 100 K but no FUV irradiation depends considerably on the geometry and the bolometric luminosity of the source. Models with a geometrically thin inner part of the envelope (for example a large outflow opening angle) have considerably less material with temperatures above 100 K but no FUV irradiation.

\section{Chemical Abundances} \label{sec:chemabu}

Chemical abundances, their evolution, differences between physical models and the impact of X-rays are studied in this section. We first study fractional abundances $n({\rm X},r')/n_{\rm tot}(r')$, with $n({\rm X},r')$ and $n_{\rm tot}(r')$, the density of the species and the total hydrogen density ($n_{\rm tot} = 2 n({\rm H}_2) + n({\rm H})$) depending on the position. The fractional abundance gives an overview of the spatial variation of the molecule. Next, the volume averaged fractional abundance is discussed. The volume averaged abundance of a species X within a radius $r$ is defined by  
\begin{equation} \label{eq:volavg}
\langle X \rangle_r = \left. \int_{\left|r'\right| < r} n({\rm X},r') dV \middle/ \int_{\left|r'\right| < r} n_{\rm tot}(r') dV \right. \ .
\end{equation}
This quantity is more related to observations, as the volume averaged abundance times the mass gives the total amount of molecules. The total amount of molecules scales with the line flux in the optically thin case, neglecting excitation effects.

In paper I, the dust and gas temperature are assumed to be equal, and evaporation is approximated by an increased initial abundance for certain temperature ranges. Here, we calculate the dust and gas temperature explicitly. Since the abundances of the species considered here do not change much in the temperature range $70 < T < 100$ K, but much more with water evaporation at 100 K, we use $T=\max(T_{\rm dust},T_{\rm gas})$ to obtain the abundances from the grid of chemical models. This choice affects only the innermost region with dust temperatures above 100 K heated by infrared photons (Figure \ref{fig:abus_std1} and Section \ref{sec:physstruc}). Abundances outside this region are found to agree to within 20\% if the gas temperature were used instead. This also shows that the abundances of the diatomic hydrides are not sensitively depending on the assumed water evaporation temperature. The evaporation temperature of the diatomic hydrides is not well known. However, as their abundance on dust grains likely is small (e.g. \citealt{Hasegawa93}), the effect of direct evaporation of diatomic hydrides to the gas phase is neglected here.

\subsection{Standard and spherically symmetric model} \label{sec:chemstd}

\begin{figure*}[tbh]
  \centering
  \includegraphics[width=1.0\hsize]{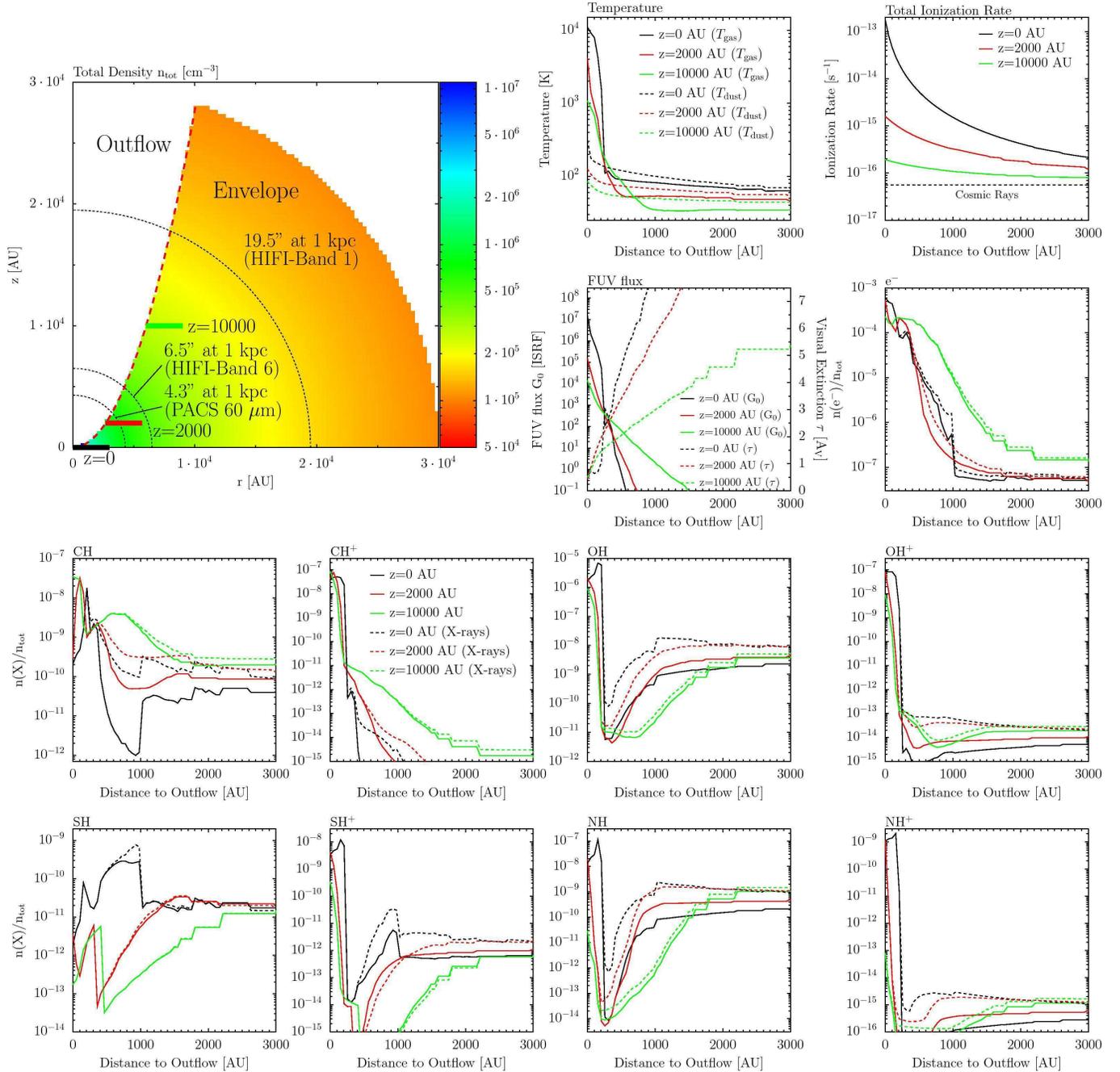}
  \caption{Abundances along cuts of constant $z$ through the model \textit{Standard}. Abundances assuming no protostellar X-ray radiation (solid line) and an X-ray luminosity of $10^{32}$ erg s$^{-1}$ (dashed line) are shown (bottom panels). The density structure and dust/gas temperature, total ionization rate and FUV flux/attenuation along cuts with $z=0, 2000, 10000$ AU are given too (top panels). The standard cosmic-ray ionization rate of $5.6 \times 10^{-17}$ s$^{-1}$ is indicated by a dashed line.}
   \label{fig:abus_std1}
\end{figure*}

\begin{figure*}[tbh]
  \centering
  \includegraphics[width=1.0\hsize]{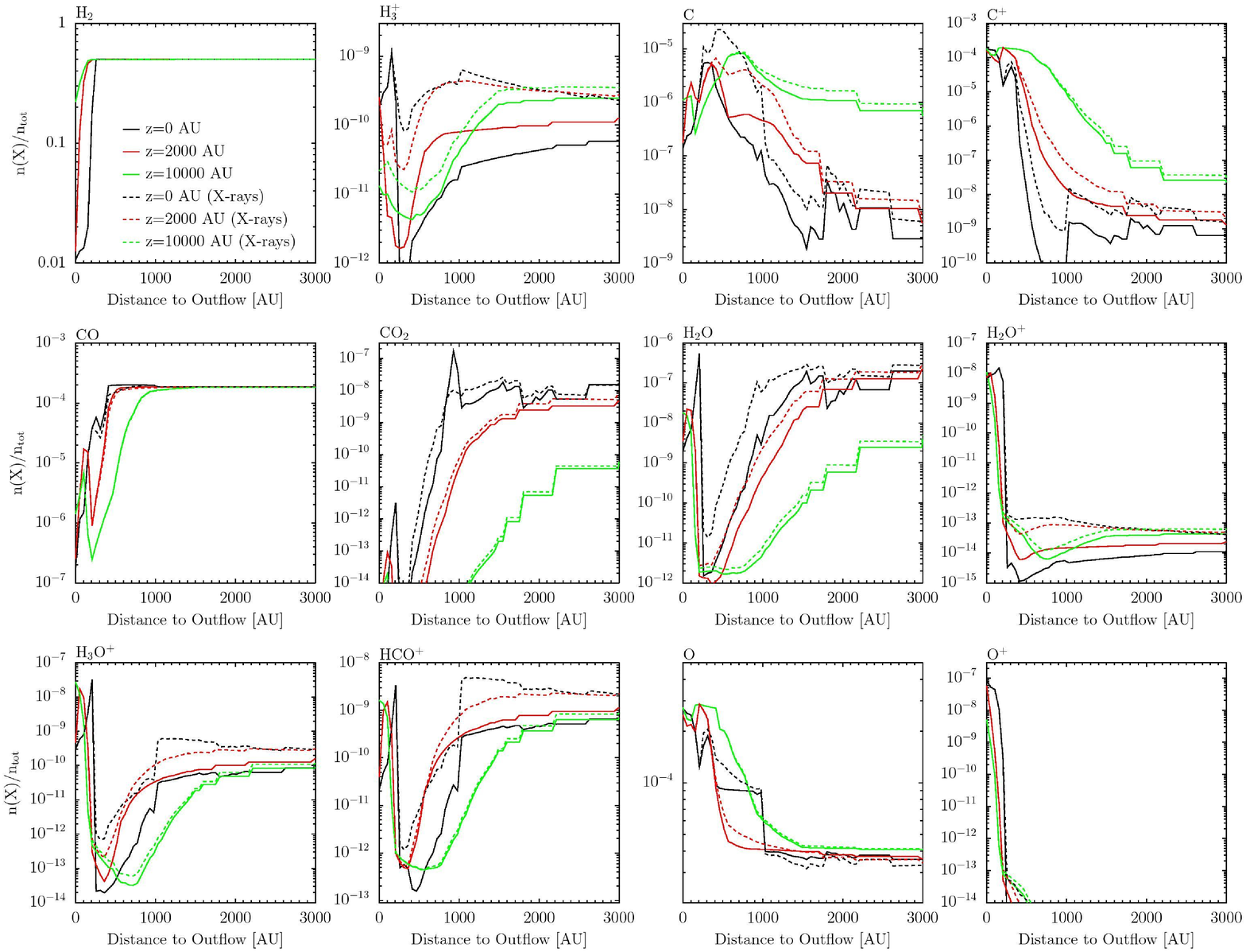}
  \caption{Abundances along cuts of constant $z$ through the model \textit{Standard}. Abundances assuming no protostellar X-ray radiation (solid line) and an X-ray luminosity of $10^{32}$ erg s$^{-1}$ (dashed line) are shown.}
   \label{fig:abus_std2}
\end{figure*}

The abundances of diatomic hydrides, related species and main molecules/atoms (H$_2$, CO, CO$_2$, H$_2$O, O, C, C$^+$) for the model \textit{Standard} are given along cuts of constant $z$ (parallel to the midplane, at $0$, $2000$ and $10000$ AU) in Figures \ref{fig:abus_std1} and \ref{fig:abus_std2}. Presented abundances are for a chemical age of \tto{5}{4} yrs (\citealt{Staeuber05}) and with a protostellar X-ray luminosity of $10^{32}$ erg s$^{-1}$ or no protostellar X-rays. The density structure of the model along with the dust and gas temperature, total ionization rate (see paper I), the FUV flux and attenuation are given in the top panels of Figure \ref{fig:abus_std1}.

As a rough classification of the abundance profiles we can divide the species into broad groups of molecules which are (i) enhanced in the outflow wall compared to the envelopes (e.g. electrons, C$^+$, CH$^+$) with different widths of the enhanced region; (ii) destroyed along the outflow (e.g. H$_2$, CO or H$_2$O); (iii) enhanced in a thin top layer and destroyed at intermediate depths (e.g. OH or H$_3$O$^+$). Why are some molecules enhanced or destroyed in different layers? The overall chemistry of FUV/X-ray irradiated molecular gas has been studied in various papers (\citealt{Hollenbach99} for a review) and for envelopes of YSOs in particular by \citet{Staeuber04,Staeuber05}. Here, we focus on the chemistry of diatomic hydrides and related species in the physical regime of the outflow walls.

\subsubsection{The chemistry in the outflow walls} \label{sec:chemflowwall}

In the top layer of the outflow wall with $\tau_V$ smaller than a few, the temperature is high, with $T_{\rm gas} > 100$ K. Due to self-shielding, H$_2$ and CO are only destroyed in a thin layer. Photoprocesses proceeding through continuum absorption, as for example the photoionization of carbon, require a minimum photon energy ($1102$\AA). Due to the higher dust opacity at shorter wavelengths (reddening), photons with higher energy are absorbed more quickly and carbon photoionization is most active at the surface. Ionized carbon forms CH$^+$ by, C$^+$ + H$_2$ $\rightarrow$ CH$^+$ + H, but this reaction is highly endoergic by 4640 K. The reaction of CH$^+$ with H$_2$ produces CH$_2^+$ and CH$_3^+$ which recombine with electrons to CH. Since the oxygen ionization energy is slightly higher than 13.6 eV, O$^+$ is produced by charge exchange with protons, O + H$^+$ $\rightarrow$ O$^+$ + H, with H$^+$ from the photodissociation of CH$^+$. Ionized hydroxyl (OH$^+$) is then formed by H$_2$ + O$^+$ $\rightarrow$ OH$^+$ + H. Subsequent reactions with H$_2$ lead from OH$^+$ to H$_2$O$^+$ and then H$_3$O$^+$, which recombines to OH and H$_2$O. Hydroxyl is however mainly formed by the endoergic reaction H$_2$ + O + 3150 K $\rightarrow$ OH + H. Water also forms by an endoergic reaction, H$_2$ + OH + 1740 K $\rightarrow$ H$_2$O + H and is photodissociated to OH + H. Reactions of OH with O and CO form O$_2$ and CO$_2$, respectively, which are both quickly photodissociated. Sulfur hydride and its ion are formed in the endoergic reactions S + H$_2$ + 6130 K $\rightarrow$ SH + H and S$^+$ + H$_2$ + 9860 K $\rightarrow$ SH$^+$ + H with S$^+$ from direct photoionization of S.  Finally, NH is created by H$_2$ + N + 18100 K $\rightarrow$ NH + H. Since the ionization potential of atomic nitrogen is higher than 13.6 eV, NH$^+$ is only produced by photoionization of NH or charge exchange H$^+$ + NH $\rightarrow$ NH$^+$ + H.

In an intermediate layer, at higher optical depth, the temperature drops below 100 K (Figure \ref{fig:abus_std1}). In this colder environment, the photodissociation of OH is still considerable but the endoergic formation reaction of H$_2$ + O $\rightarrow$ OH + H is no longer efficient. This explains the drop of the OH abundance in this layer. The drop is also seen in other molecules (SH$^+$, NH, NH$^+$, H$_2$O; group (iii) in Section \ref{sec:chemstd}). For H$_3$O$^+$, the situation is different: Since CH$^+$ is not available in this layer, the main production of H$^+$ is disabled and subsequently, no O$^+$ is available. Thus, the abundances of OH$^+$, H$_2$O$^+$ and H$_3$O$^+$ are decreased.

We conclude that the concentration of many outflow wall enhanced species towards the outflow is a consequence of (atomic) ionization being only active at the surface due to a quick shielding of short wavelength photons by dust (C$^+$ and S$^+$) combined with the high required temperature to overcome the activation energy for the formation  (especially CH$^+$, OH, SH, SH$^+$, NH and NH$^+$).

In the midplane ($z=0$), at a distance of about 1000 AU to the protostar, the influence of H$_2$S, H$_2$O and CO$_2$ evaporation can be seen. At this point, FUV radiation is present but too weak to heat the gas above 100 K, while IR radiation still heats the dust grains to $T_{\rm dust} > 100$ K. The evaporated water is quickly photodissociated and its abundance is not increased. The higher amount of sulphur in the gas phase leads to an enhancement of SH and SH$^+$ which can been seen. Also affected are C$^+$ and CH, which have an increased abundance in colder regions due to reaction partners being less abundant. C$^+$ is destroyed in reactions with atomic sulphur, while CH is destroyed with atomic oxygen. Among the species discussed here (SH, SH$^+$, CH and CH$^+$), only SH reaches abundances in this region that are higher compared to the outflow walls. Thus, SH emission may be dominated by this region with sulphur evaporated to the gas phase (Section \ref{sec:compmods} and \ref{sec:lineflux}).

We next consider HCO$^+$ which is predicted to be an ionization tracer (e.g. \citealt{vdTak00b} or \citealt{Savage04b}). In the top layer of the outflow wall, it is formed by the reaction of H$_2$ with HOC$^+$ and CO$^+$. The huge enhancement of CO$^+$ by four orders of magnitude in a thin layer along the outflow has been discussed in paper II. Going deeper into the outflow-wall, HCO$^+$ is destroyed by electron recombination and then enhanced by the reaction of CO with H$_3^+$. The H$_3^+$ ion is a key species for the X-ray driven chemistry (\citealt{Maloney96} or \citealt{Staeuber05}) and formed by cosmic-ray or X-ray ionization of H$_2$ followed by a quick reaction with H$_2$. The high abundance of H$_3^+$ at the surface of the outflow wall is maintained by the reaction NH$^+$ + H$_2$ $\rightarrow$ N + H$_3^+$. Deeper in the outflow wall, the high electron fraction leads to quick electron recombination of H$_3^+$.

An important difference between X-ray and FUV driven chemistry is the electron fraction. For fluxes of 1 erg s$^{-1}$ cm$^{-2}$ ($G_0 \approx 1000$ ISRF) the electron fraction at a total density of $10^6$ cm$^{-3}$ is approximately $x(e^-)=n(e^-)/n_{\rm tot}  \sim 10^{-4}$ for FUV but only $10^{-6}$ for X-rays. Since the luminosity in the FUV band is 4 orders of magnitude higher than the X-ray luminosity, the large electron fraction is mainly produced by FUV radiation. Since this electron fraction leads to the destruction of H$_3^+$, the influence of X-rays is dwarfed by FUV radiation. In addition, many molecules predicted to be X-ray tracers by \citet{Staeuber05}, are much more enhanced by FUV compared to X-rays. Particular examples are CH$^+$ and OH$^+$, which both have fractional abundances below $10^{-12}$ in the spherical model but increase to more than $10^{-8}$ in the outflow walls. We conclude that the effect of X-rays compared to FUV on the chemistry is small in a geometry which allows the much stronger FUV radiation to escape from the innermost part.

\subsubsection{Volume averaged abundances} \label{sec:volflowwall}

How is the total amount of a molecule/atom altered by the enhancement or absence in the outflow wall? For example a thin layer of strongly enhanced CO$^+$ is found in paper II to enhance the total amount by three orders of magnitude in agreement with observations. We note that CO$^+$ is related to the diatomic  hydrides studied here through formation by C$^+$ + OH $\rightarrow$ CO$^+$ + H. FUV destroyed species on the other hand, should not decrease their total abundance significantly since only a relatively small fraction of the envelope mass is cut out by a cavity (Section \ref{sec:physstruc}). In Table \ref{tab:avgstd1d2d}, we give volume averaged abundances within a radius of 4000 and 20000 AU for the model \textit{Standard} and compare it to the spherical model. In the following we consider a deviation larger than a factor of three to be significant, since similar models (\citealt{Doty04}) qualify a factorial deviation of this factor between a model and observation as a good agreement.

\begin{table*}[tbh]
\footnotesize
\caption{Volume averaged abundances $\langle X \rangle_r$ of selected species for the model \textit{Standard} (``2D'') and the spherically symmetric model (``1D''). Abundances averaged over a radius of 4000 and 20000 AU from the protostar and assuming different chemical ages and protostellar X-ray luminosities are given. The ratio of averaged abundances between the model \textit{Standard} and the spherically symmetric model is given on the right.\label{tab:avgstd1d2d}}
\centering
\begin{tabular}{l|ccccc|ccc|ccc}
\tableline
Model      & 2D        & 2D        & 2D        & 2D        & 2D        & 1D        & 1D        & 1D      & \multicolumn{3}{c}{Ratio 2D/1D} \\
\tableline
X-rays [erg s$^{-1}$]     & 1(32)     & 1(32)     & -         & 1(32)     & 1(32)     & 1(32)     & 1(32)     & -$^a$   & 1(32) & 1(32) & - \\ 
Chem. age [yrs]  & 5(4)      & 5(4)      & 5(4)      & 3(3)      & 3(5)      & 5(4)      & 5(4)      & 5(4)    & 5(4) & 5(4) & 5(4) \\
Averaged [AU]    & 20000     & 4000      & 4000      & 20000     & 20000     & 20000     & 4000      & 4000    & 20000 & 4000 & 4000 \\
\tableline
H$_2$            & 5(-1)     & 5(-1)     & 5(-1)     & 5(-1)     & 5(-1)     & 5(-1)     & 5(-1)     & 5(-1)     & 1.0       & 0.9       & 0.9       \\
e$^-$            & 8(-6)     & 5(-5)     & 5(-5)     & 6(-6)     & 8(-6)     & 1(-8)     & 2(-8)     & 6(-9)     & \textbf{8(2)}      & \textbf{3(3)}      & \textbf{8(3)}      \\
H$_3^+$          & 3(-10)    & 3(-10)    & 5(-11)    & 3(-10)    & 3(-10)    & 3(-10)    & 4(-10)    & 6(-11)    & 1.0       & 0.7       & 0.8       \\
CH               & 6(-10)    & 2(-9)     & 2(-9)     & 9(-10)    & 6(-10)    & 5(-11)    & 8(-11)    & 2(-11)    & \textbf{11}        & \textbf{30}        & \textbf{1(2)}      \\
CH$^+$           & 4(-10)    & 4(-9)     & 4(-9)     & 4(-10)    & 4(-10)    & 4(-16)    & 6(-16)    & 6(-17)    & \textbf{1(6)}      & \textbf{8(6)}      & \textbf{7(7)}      \\
OH               & 2(-8)     & 1(-7)     & 1(-7)     & 2(-8)     & 2(-8)     & 9(-9)     & 1(-8)     & 3(-9)     & 1.9       & \textbf{9.5}       & \textbf{43}        \\
OH$^+$           & 1(-10)    & 3(-9)     & 3(-9)     & 1(-10)    & 1(-10)    & 2(-14)    & 3(-14)    & 4(-15)    & \textbf{6(3)}      & \textbf{9(4)}      & \textbf{7(5)}      \\
SH               & 3(-11)    & 2(-11)    & 2(-11)    & 2(-10)    & 2(-11)    & 2(-10)    & 1(-9)     & 3(-9)     & \textbf{0.1}       & \textbf{0.01}      & \textbf{6(-3)}     \\
SH$^+$           & 1(-11)    & 2(-10)    & 2(-10)    & 1(-11)    & 1(-11)    & 9(-12)    & 9(-11)    & 2(-12)    & 1.1       & 2.1       & \textbf{1(2)}      \\
NH               & 1(-9)     & 2(-9)     & 9(-10)    & 1(-9)     & 1(-9)     & 5(-10)    & 7(-10)    & 6(-11)    & 2.4       & 2.3       & \textbf{15}        \\
NH$^+$           & 1(-12)    & 3(-11)    & 3(-11)    & 1(-12)    & 1(-12)    & 1(-15)    & 2(-15)    & 3(-16)    & \textbf{8(2)}      & \textbf{2(4)}      & \textbf{1(5)}      \\
C$^+$            & 7(-6)     & 3(-5)     & 3(-5)     & 4(-6)     & 7(-6)     & 4(-10)    & 7(-10)    & 1(-10)    & \textbf{2(4)}      & \textbf{4(4)}      & \textbf{2(5)}      \\
C                & 5(-7)     & 2(-6)     & 7(-7)     & 1(-7)     & 6(-7)     & 4(-8)     & 3(-8)     & 1(-8)     & \textbf{11}        & \textbf{70}        & \textbf{51}        \\
CO               & 2(-4)     & 1(-4)     & 1(-4)     & 2(-4)     & 2(-4)     & 2(-4)     & 2(-4)     & 2(-4)     & 1.0       & 0.8       & 0.8       \\
CO$_2$           & 5(-9)     & 4(-9)     & 5(-9)     & 1(-9)     & 1(-8)     & 3(-8)     & 7(-8)     & 9(-7)     & \textbf{0.2}       & \textbf{0.06}      & \textbf{5(-3)}     \\
H$_2$O           & 2(-7)     & 9(-8)     & 6(-8)     & 5(-8)     & 3(-7)     & 8(-7)     & 6(-7)     & 5(-6)     & \textbf{0.3}       & \textbf{0.1}       & \textbf{0.01}      \\
H$_2$O$^+$       & 5(-11)    & 5(-10)    & 5(-10)    & 5(-11)    & 5(-11)    & 4(-14)    & 7(-14)    & 1(-14)    & \textbf{1(3)}      & \textbf{7(3)}      & \textbf{5(4)}      \\
H$_3$O$^+$       & 4(-10)    & 1(-9)     & 9(-10)    & 3(-10)    & 4(-10)    & 1(-9)     & 2(-9)     & 6(-10)    & 0.4       & 0.6       & 1.5       \\
HCO$^+$          & 2(-9)     & 1(-9)     & 4(-10)    & 2(-9)     & 2(-9)     & 4(-9)     & 8(-9)     & 2(-9)     & 0.5       & \textbf{0.2}       & \textbf{0.2}       \\
O                & 5(-5)     & 8(-5)     & 8(-5)     & 5(-5)     & 4(-5)     & 4(-5)     & 3(-5)     & 3(-5)     & 1.3       & 2.5       & 2.4       \\
\tableline
\end{tabular}
\begin{flushleft}
\footnotesize{Note: $a(b)=a \times 10^b$.\\
$^a$FUV also switched off.}
\end{flushleft}
\end{table*}

Outflow wall enhanced species indeed have a significantly enhanced volume averaged abundance compared to the spherically symmetric model. Most enhanced are the ionized molecules C$^+$, CH$^+$, OH$^+$, NH$^+$, H$_2$O$^+$ and the electron fraction with an enhancement larger than 3 orders of magnitude and up to almost 7 orders of magnitude. On the other hand, it is found that the FUV destruction of H$_2$O, CO$_2$ and SH reduces the volume averaged abundance less, by at most a factor of 100. Both enhancement and reduction are seen more in the abundance averaged over 4000 AU rather than 20000 AU. The width of the region with FUV enhancement is important too. For example the volume averaged abundance of SH$^+$ is only increased by a factor of about two despite the enhancement by about 2 orders of magnitude in a very thin layer on the surface of the outflow wall. Similarly, the abundance of NH is only increased in the inner part of the envelope due to the strong temperature dependence of the abundance and the volume averaged abundance is only increased by a factor of two. This strong temperature dependence and consequently the concentration to the inner part also explains the largest difference of enhancement of NH$^+$ between the volume averages of 4000 AU and 20000 AU.

The temporal evolution is studied in Table \ref{tab:avgstd1d2d} by the volume averaged abundances within 20000 AU for a chemical age of $10^{11}$ s (\tto{3}{3} yrs) and $10^{13}$ s (\tto{3}{5} yrs), and compared to the chemical age of \tto{5}{4} yrs found by \citet{Staeuber05}. The abundances averaged over 4000 AU are not shown since they are less time dependent. The outflow wall enhanced species have very short chemical time-scale for their formation and thus do not show any temporal evolution. We find that CH$^+$ in the enhanced region only requires of order 10 yrs to reach the final fractional abundance to within a factor of two. The time-scales of OH$^+$, NH$^+$ and H$_2$O$^+$ are similarly short. Assuming an upper limit of 10 km s$^{-1}$ for the Alfv\'en and sound velocity, the irradiated gas would move up to a distance of about 25 AU within the chemical timescale which justifies our approach to ignore the dynamical evolution of the gas. This is further justified by the fact that diatomic hydrides produced in the outflow walls diffusing into the cloud are quickly destroyed by reactions with other species. To test this, a chemical model presented in paper I has been run for conditions of the cold and not irradiated gas but with initially increased abundances of the diatomic hydrides. Since the C$^+$ enhanced region is thicker, contributions from colder and less irradiated regions with longer timescales are important and the volume averaged abundance shows some temporal evolution (within a factor of two).

\subsection{Comparing different geometries} \label{sec:compmods}

The influence of different physical structures (Section \ref{sec:physstruc}) on chemical abundances is studied in this section. Tables \ref{tab:avgmods1} and \ref{tab:avgmods2} give the volume averaged abundances of molecules studied in the previous section. A chemical age of \tto{5}{4} yrs and an X-ray luminosity of $L_{\rm X} = 10^{32}$ erg s$^{-1}$ are used. The abundance is averaged over 4000 AU (all models) and 20000 AU (selected models). To obtain the total amount of a molecule, the mass of the region over which is averaged ($M_{4000}$ and $M_{20000}$) is given relative to the model \textit{Standard}. The molecules which were found in Section \ref{sec:volflowwall} to be significantly enhanced/destroyed or unchanged are marked with ``+''/``-'' or ``0'', respectively.

\begin{table*}[tbh]
\tablewidth{0pt}
\footnotesize
\caption{Ratios of the volume averaged abundances (Eq. \ref{eq:volavg}) to the model \textit{Standard}. A chemical age of \tto{5}{4} yrs and a protostellar X-ray luminosity of $10^{32}$ erg s$^{-1}$ are assumed. The mass over which is averaged ($M_{4000}$ and $M_{20000}$) is given relative to model \textit{Standard}. \label{tab:avgmods1}}
\centering
\begin{tabular}{lll|cccccccccc}
\tableline
 & Model                 & & H$_2$            & e$^-$            & H$_3^+$          & CH               & CH$^+$           & OH               & OH$^+$           & SH               & SH$^+$           & NH \\
\tableline
\multicolumn{2}{l}{\textbf{Within 4000 AU}} & $M_{4000}$ & \textbf{0}$^a$ & \textbf{+} & \textbf{0} & \textbf{(+)} & \textbf{+} & \textbf{(+)} & \textbf{+} & \textbf{-} & \textbf{+} & \textbf{0}  \\
  & \textit{Spherical}    &  2.1   &      1.1 &   \textbf{4(-4)} &              1.4 &   \textbf{3(-2)} &   \textbf{1(-7)} &     \textbf{0.1} &   \textbf{1(-5)} &    \textbf{1(2)} &              0.5 &              0.4  \\  
  & \textit{Standard}     &  1     &      1   &              1   &              1   &              1   &              1   &              1   &              1   &              1   &              1   &              1    \\  
  & \textit{Disk}         &  9.1   &      1.1 &   \textbf{1(-1)} &   \textbf{3(-2)} &     \textbf{0.2} &   \textbf{7(-2)} &              0.7 &   \textbf{6(-2)} &              2.5 &   \textbf{8(-2)} &   \textbf{5(-2)}  \\ 
  & \textit{$\alpha$20}   &  1.8   &      1.0 &     \textbf{0.2} &              1.1 &     \textbf{0.2} &     \textbf{0.2} &              0.9 &     \textbf{0.3} &    \textbf{1(2)} &     \textbf{0.3} &              1.1  \\
  & \textit{$\alpha$40}   &  1.5   &      1.0 &              0.5 &              1.2 &              0.4 &              0.5 &              0.5 &              0.6 &     \textbf{4.0} &              0.6 &              1.1  \\ 
  & \textit{$\alpha$80}   &  0.64  &      1.0 &              1.5 &              0.7 &              2.9 &              1.8 &              1.9 &              1.2 &              0.4 &              1.6 &              1.0  \\ 
  & \textit{$\alpha$100}  &  0.36  &      0.9 &              2.2 &              0.4 &     \textbf{8.0} &              2.9 &     \textbf{3.8} &              1.5 &     \textbf{0.1} &              2.3 &              1.2  \\ 
  & \textit{$\gamma$0}    &  1     &      1.0 &              1.0 &              1.0 &              0.9 &              1.0 &              0.8 &              0.9 &              0.9 &              0.8 &              0.9  \\ 
  & \textit{$\gamma$0.01} &  1     &      1.0 &              0.8 &              1.0 &              0.9 &              0.8 &              0.8 &              0.5 &              1.4 &              0.9 &              1.0  \\ 
  & \textit{$\gamma$0.1}  &  1     &      1.1 &   \textbf{4(-2)} &              1.6 &   \textbf{9(-2)} &   \textbf{9(-3)} &     \textbf{0.2} &   \textbf{1(-2)} &     \textbf{4.0} &     \textbf{0.3} &              1.8  \\ 
  & \textit{b1.5}         &  1.3   &      1.0 &              0.6 &              1.0 &              0.6 &              0.6 &              0.8 &              0.6 &    \textbf{3(1)} &              0.7 &              1.4  \\ 
  & \textit{b2.5}         &  0.76  &      1.0 &              1.4 &              1.0 &              1.7 &              1.6 &              1.4 &              1.4 &              0.6 &              1.4 &              1.0  \\ 
  & \textit{L2e3}         &  1     &      1.0 &              0.5 &              1.2 &              1.7 &              0.5 &              1.1 &     \textbf{0.1} &              0.8 &              0.6 &              1.2  \\ 
  & \textit{L1e4}         &  1     &      1.0 &              0.8 &              1.1 &              1.3 &              0.9 &              1.1 &              0.8 &              0.7 &              1.0 &              1.3  \\ 
  & \textit{L4e4}         &  1     &      1.0 &              1.3 &              0.9 &              0.9 &              1.2 &              0.8 &              1.0 &              2.7 &              0.9 &              0.7  \\
\tableline
\multicolumn{2}{l}{\textbf{Within 20000 AU}} & $M_{20000}$ &  \\
  & \textit{Spherical}    &  1.8   &      1.0 &   \textbf{1(-3)} &              1.0 &   \textbf{9(-2)} &   \textbf{9(-7)} &              0.5 &   \textbf{2(-4)} &     \textbf{6.8} &              0.9 &              0.4  \\ 
  & \textit{Standard}     &  1     &      1   &              1   &              1   &              1   &              1   &              1   &              1   &              1   &              1   &              1    \\ 
  & \textit{Disk}         &  1.2   &      1.0 &              0.8 &              0.8 &              0.9 &              0.7 &              1.6 &              0.5 &              1.0 &              0.6 &              0.8  \\ 
  & \textit{$\alpha$20}   &  1.2   &      1.0 &     \textbf{0.2} &              1.0 &              0.4 &     \textbf{0.2} &              0.8 &     \textbf{0.3} &     \textbf{5.7} &              0.6 &              0.7  \\ 
  & \textit{$\gamma$0.1}  &  1     &      1.0 &   \textbf{1(-2)} &              1.1 &     \textbf{0.3} &   \textbf{2(-3)} &              0.7 &   \textbf{8(-3)} &              1.2 &              0.4 &              1.1  \\ 
  & \textit{L2e3}         &  1     &      1.0 &              0.5 &              1.0 &              1.2 &     \textbf{0.2} &              0.8 &   \textbf{8(-2)} &              1.3 &              0.5 &              1.0  \\ 
\tableline
\end{tabular}
\begin{flushleft}
\footnotesize{Note: $a(b)=a \times 10^b$.\\
$^a$Ratio to the spherical model within 4000 AU (Table \ref{tab:avgstd1d2d}; \textbf{+} more than $10^3$, \textbf{-} less than 0.1, \textbf{(+)} more than 3, \textbf{0} within 0.1 to 3).}
\end{flushleft}
\end{table*}

\begin{table*}[tbh]
\tablewidth{0pt}
\footnotesize
\caption{Ratios of the volume averaged abundances (Eq. \ref{eq:volavg}) to the model \textit{Standard}. A chemical age of \tto{5}{4} yrs and a protostellar X-ray luminosity of $10^{32}$ erg s$^{-1}$ are assumed. The mass over which is averaged ($M_{4000}$ and $M_{20000}$) is given relative to model \textit{Standard}. \label{tab:avgmods2}}
\centering
\begin{tabular}{lll|cccccccccc}
\tableline  
  &                       & & NH$^+$           & C$^+$           & C &  CO               & CO$_2$           & H$_2$O           & H$_2$O$^+$       & H$_3$O$^+$       & HCO$^+$          & O                 \\
\tableline
\multicolumn{2}{l}{\textbf{Within 4000 AU}} & $M_{4000}$ &  \textbf{+}$^a$ & \textbf{+} & \textbf{(+)} & \textbf{0} & \textbf{-} & \textbf{-} & \textbf{+} & \textbf{0} & \textbf{0} & \textbf{0}  \\
  & \textit{Spherical}    &  2.1   &        \textbf{6(-5)} &   \textbf{3(-5)}&   \textbf{2(-2)} & 1.3 &    \textbf{2(1)} &     \textbf{6.8} &   \textbf{1(-4)} &              1.6 &     \textbf{5.9} &              0.4       \\
  & \textit{Standard}     &  1     &                   1   &              1  &              1   & 1   &              1   &              1   &              1   &              1   &              1   &              1         \\ 
  & \textit{Disk}         &  9.1   &        \textbf{8(-2)} &     \textbf{0.1}&     \textbf{0.2} & 1.3 &    \textbf{3(2)} &    \textbf{7(1)} &   \textbf{8(-2)} &   \textbf{9(-2)} &   \textbf{6(-2)} &              0.5       \\
  & \textit{$\alpha$20}   &  1.8   &          \textbf{0.3} &     \textbf{0.2}&              0.4 & 1.3 &     \textbf{4.7} &     \textbf{4.7} &     \textbf{0.2} &              0.4 &              1.5 &              0.6       \\
  & \textit{$\alpha$40}   &  1.5   &                   0.7 &              0.5&              0.8 & 1.2 &     \textbf{3.0} &              2.5 &              0.4 &              0.6 &              1.5 &              0.7       \\
  & \textit{$\alpha$80}   &  0.64  &                   1.2 &              1.6&              1.1 & 0.8 &   \textbf{4(-2)} &     \textbf{0.1} &              1.6 &              1.7 &              0.5 &              1.3       \\
  & \textit{$\alpha$100}  &  0.36  &                   1.6 &              2.4&              1.3 & 0.5 &   \textbf{2(-4)} &     \textbf{0.2} &              2.4 &              2.8 &              0.5 &              1.8       \\
  & \textit{$\gamma$0}    &  1     &                   0.7 &              1.0&              1.0 & 1.0 &              1.2 &              1.2 &              0.9 &              1.0 &              1.0 &              1.0       \\
  & \textit{$\gamma$0.01} &  1     &                   0.6 &              1.0&              1.0 & 1.0 &              1.3 &              1.1 &              0.9 &              0.9 &              1.1 &              1.0       \\
  & \textit{$\gamma$0.1}  &  1     &          \textbf{0.1} &   \textbf{3(-2)}&              0.9 & 1.3 &     \textbf{3.2} &              2.5 &   \textbf{2(-2)} &              0.4 &              2.0 &              0.5       \\
  & \textit{b1.5}         &  1.3   &                   0.6 &              0.6&              0.7 & 1.1 &              2.5 &              2.3 &              0.6 &              0.6 &              1.3 &              0.8       \\
  & \textit{b2.5}         &  0.76  &                   1.3 &              1.4&              1.2 & 0.9 &     \textbf{0.2} &     \textbf{0.3} &              1.6 &              1.4 &              0.7 &              1.2       \\
  & \textit{L2e3}         &  1     &          \textbf{0.2} &              0.7&              0.9 & 1.1 &              0.7 &              1.6 &              0.6 &              1.0 &              1.1 &              0.8       \\
  & \textit{L1e4}         &  1     &                   0.9 &              1.0&              0.9 & 1.0 &              1.0 &              1.3 &              1.0 &              0.8 &              1.1 &              0.9       \\
  & \textit{L4e4}         &  1     &                   0.8 &              1.1&              1.2 & 1.0 &              1.4 &              1.0 &              1.2 &              1.1 &              0.9 &              1.1       \\
\tableline
\multicolumn{2}{l}{\textbf{Within 20000 AU}}& $M_{20000}$ \\
  & \textit{Spherical}    &  1.8   &       \textbf{1(-3)} &   \textbf{7(-5)}&   \textbf{9(-2)} & 1.0 &      \textbf{6.3} &     \textbf{3.5} &   \textbf{1(-3)} &              2.4 &              2.2 &              0.8       \\
  & \textit{Standard}     &  1     &                  1   &              1  &              1   & 1   &               1   &              1   &              1   &              1   &              1   &              1         \\ 
  & \textit{Disk}         &  1.2   &                  0.5 &              0.8&              0.9 & 1.0 &     \textbf{5(1)} &     \textbf{6.1} &              0.7 &              0.7 &              0.7 &              1.0       \\
  & \textit{$\alpha$20}   &  1.2   &                  0.4 &     \textbf{0.2}&              0.4 & 1.0 &      \textbf{3.3} &              1.8 &     \textbf{0.2} &              1.7 &              1.8 &              0.8       \\
  & \textit{$\gamma$0.1}  &  1     &         \textbf{0.1} &   \textbf{4(-3)}&     \textbf{0.1} & 1.0 &               1.4 &              1.3 &   \textbf{7(-3)} &              0.9 &              1.2 &              0.8       \\
  & \textit{L2e3}         &  1     &         \textbf{0.2} &              0.5&              1.1 & 1.0 &               1.6 &              1.4 &     \textbf{0.2} &              0.9 &              1.0 &              0.9       \\
\tableline
\end{tabular}
\begin{flushleft}
\footnotesize{Note: $a(b)=a \times 10^b$.\\
$^a$Ratio to the spherical model within 4000 AU (Table \ref{tab:avgstd1d2d}; \textbf{+} more than $10^3$, \textbf{-} less than 0.1, \textbf{(+)} more than 3, \textbf{0} within 0.1 to 3).}
\end{flushleft}
\end{table*}

For the outflow wall enhanced species CH$^+$, OH$^+$, SH$^+$, NH$^+$, C$^+$, H$_2$O$^+$ and electrons, the volume averaged abundance is most changed in the models \textit{Disk}, \textit{$\alpha$20} and \textit{$\gamma$0.1}. While for \textit{$\alpha$20} and \textit{$\gamma$0.1} averaging over 4000 and 20000 AU leads to a significantly lower abundance, the differences in the \textit{Disk} model are only seen for the average over 4000 AU. However, when scaling with the higher mass of the \textit{Disk} model within 4000 AU, the total amount of these charged species remains approximately constant. This reflects that the fractional abundance of outflow wall enhanced species is approximately inversely proportional to the density (paper I). In addition, the higher density yields a lower gas temperature and thus a decreased amount of molecules which require a high temperature for formation (e.g. CH$^+$). Quantitatively, we find that the volume averaged abundances of outflow wall enhanced species behave similarly as the FUV irradiated mass, discussed in Section \ref{sec:physstruc}.

The FUV destroyed molecules SH, H$_2$O and CO$_2$ have a similar dependence of the volume averaged abundance. Like the region without FUV irradiation but temperature above $100$ K (Section \ref{sec:physstruc}), they are more dependent on the opening angle (\textit{$\alpha$20}-\textit{$\alpha$100}) and the shape (\textit{b1.5} and \textit{b2.5}) than the FUV enhanced molecules. Quantitatively, however, the dependence of the volume averaged abundance on the physical models differs more between these three species than for the outflow wall enhanced species. This is due to different effects entering the abundances of SH, H$_2$O and CO$_2$, like evaporation combined with a different depth dependence of the photodissociation rates. For example, the larger dependence of CO$_2$ on the opening angle for models \textit{$\alpha$80} and \textit{$\alpha$100} is a result of the concentration of the molecule towards the center combined with the CO$_2$ photodissociation rate dropping faster with extinction compared to H$_2$O and SH. On the other hand, SH is more enhanced in the models \textit{$\alpha$20} and \textit{b1.5} due to the factor of 10 larger increase in abundance in the hot-core region. Quantitatively, photodissociated and hot-core enhanced species thus have a more complex dependence on the geometry than FUV enhanced species.

Shocks, which are not accounted for here, may release H$_2$O and CO$_2$ to the gas phase. In a fast J-shock with $v_S = 80$ km s$^{-1}$ and a density of $10^5$ cm$^{-3}$, the width of dust grains with temperature above 100 K is of order 700 AU (\citealt{Hollenbach89}). Thus along the outflow walls, H$_2$O and CO$_2$ molecules released by shocks from the grain surface are likely photodissociated (Figure \ref{fig:abus_std1} and \ref{fig:abus_std2}). A detailed modeling of shocks along the outflow walls including the strong protostellar FUV irradiation is warranted.

\begin{figure*}[tbh]
  \centering
  \includegraphics[width=0.65\hsize]{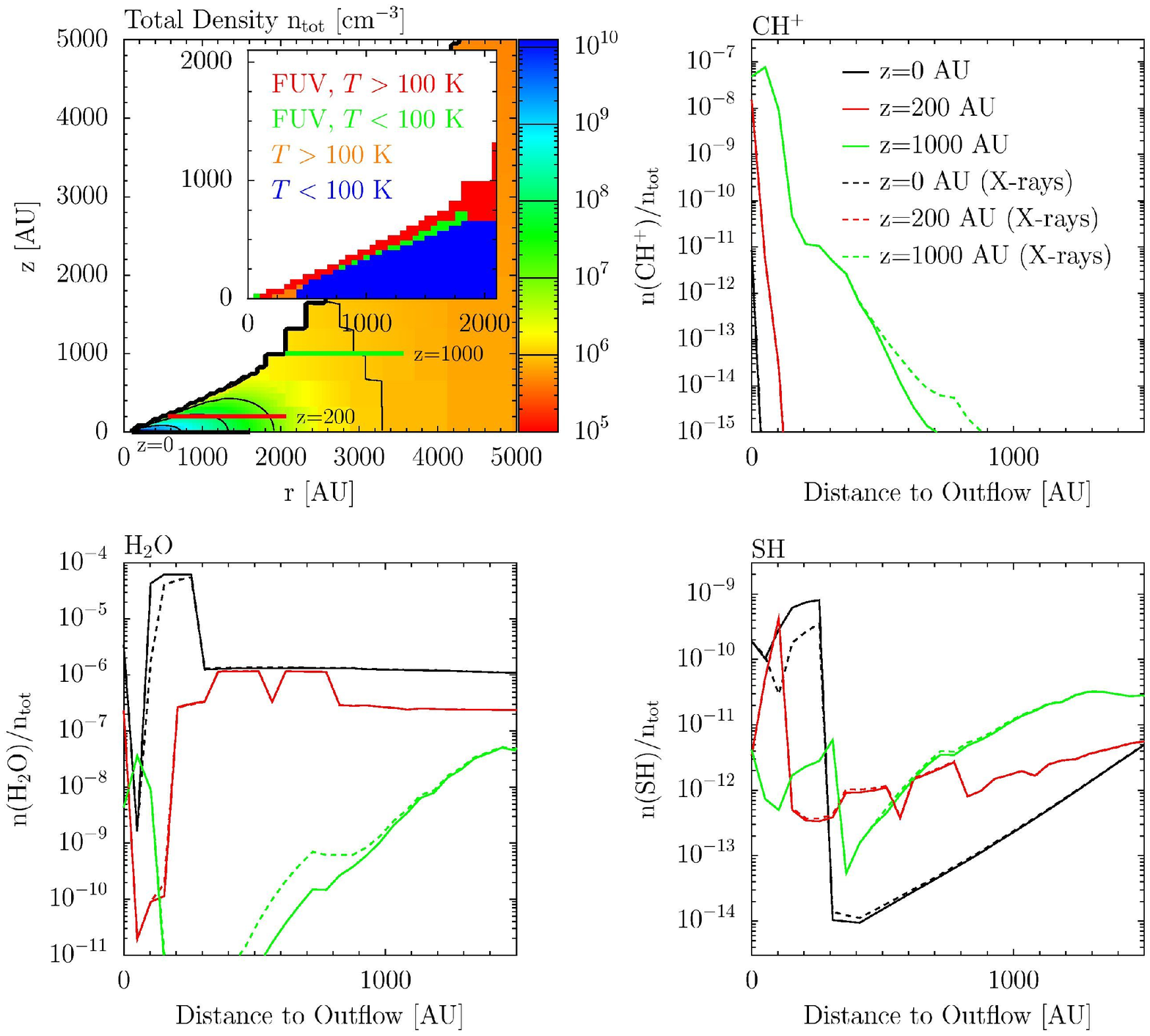}
  \caption{Density structure and regions with $T>100$ K and/or FUV irradiation of the model \textit{Disk}. Abundances of CH$^+$, SH and H$_2$O are given along cuts of constant $z$ (in the midplane and at $z=200,1000$ AU). The model with X-rays (dashed line) assumes an X-ray luminosity of $10^{32}$ erg s$^{-1}$.}
   \label{fig:disk}
\end{figure*}

Of the molecules deemed insensitive to geometry in Section \ref{sec:volflowwall} (H$_2$, H$_3^+$, NH, CO, H$_3$O$^+$, HCO$^+$ and O), the presence of a disk (Figure \ref{fig:disk}) show significant differences for H$_3^+$, NH, H$_3$O$^+$, HCO$^+$. This is again due to the combination of a lower gas temperature due to the higher density in the disk and subsequently faster recombination rate. The total amount of these molecules thus remains approximately the same as in the model \textit{Standard}.

\begin{figure*}[tbh]
  \centering
  \includegraphics[width=1.0\hsize]{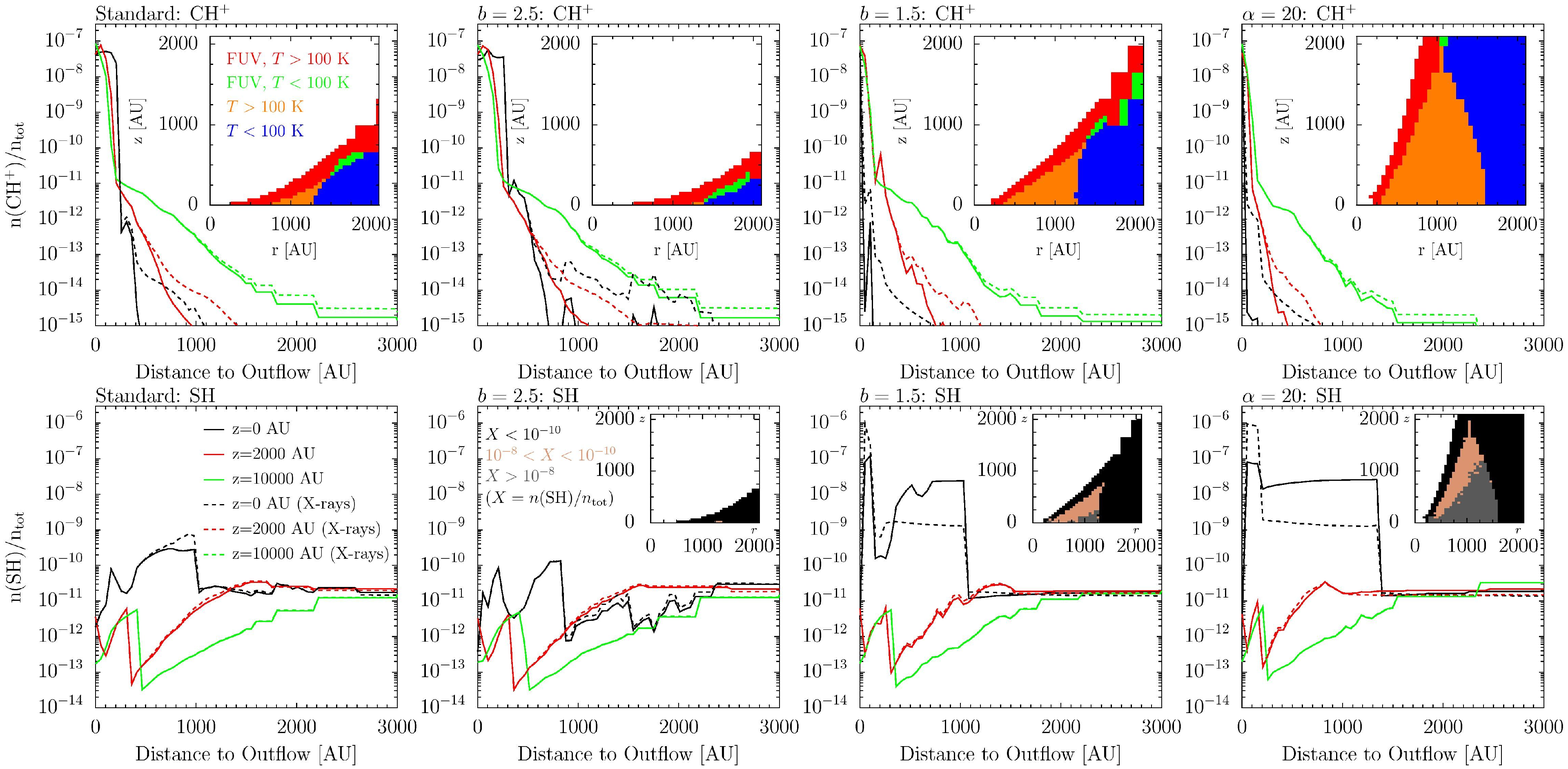}
  \caption{Abundances of CH$^+$ and SH for the models \textit{Standard}, \textit{b2.5}, \textit{b1.5} and \textit{$\alpha$20}. The abundances are given for cuts of constant $z$ assuming no protostellar X-ray radiation (solid line) and an X-ray luminosity of $10^{32}$ erg s$^{-1}$ (dashed line). As an inset to the plots for CH$^+$, the regions with FUV irradiation and/or temperature $>100$ K are given. The inset to the plots for SH show the abundance of SH divided into regions with fractional abundances $>10^{-8}$, $<10^{-10}$ and in between.}
   \label{fig:geomdiff}
\end{figure*}

Figure \ref{fig:geomdiff} shows the abundance of the FUV enhanced CH$^+$ and the FUV destroyed SH for models \textit{Standard}, \textit{b2.5}, \textit{b1.5} and \textit{$\alpha$20}. In the midplane, the models \textit{b1.5} and \textit{$\alpha$20} have a higher abundance of SH by two orders of magnitude, while the outflow wall with enhanced CH$^+$ abundance is much thinner. The region without FUV irradiation but temperature above 100 K is shown in the inset of the upper panel. This region coincides with the region having an SH fractional abundance larger than $10^{-10}$, shown as inset in the lower panel. The region with fractional abundance larger than $10^{-8}$ shows a layer structure with the molecule being more photodissociated closer to the outflow wall. At larger distances ($z=2000$ and $z=10000$), the differences between the models are smaller, as found in the previous section for the volume averaged abundances.

The abundances of SH, CH$^+$ and H$_2$O in the \textit{Disk} model are given in Figure \ref{fig:disk}. The layer with high temperature and enhanced CH$^+$ abundance (disk atmosphere) is thinner than in the outflow wall models in Figure \ref{fig:geomdiff} because of the higher density in the disk surface. Due to the larger density dependence of the SH abundance compared to H$_2$O and CO$_2$, SH does not reach a fractional abundance of $10^{-8}$ as in the models \textit{1.5} and \textit{$\alpha$20}. The water abundance on the other hand is even higher in the FUV shielded midplane of the \textit{Disk} model compared to the outflow wall models. This is also seen in Tables \ref{tab:avgmods1} and \ref{tab:avgmods2}, where the volume averaged abundances of CO$_2$ and H$_2$O are much more enhanced compared to SH.

\subsection{X-ray driven chemistry} \label{sec:xdrivchem}

X-rays affect only the innermost region due to geometrical dilution $\propto r^{-2}$ (Section \ref{sec:physstruc}). In Table \ref{tab:avgstd1d2d} we thus give the volume averaged abundance within 4000 AU. In the two-dimensional model (\textit{Standard}), the largest differences between the model for an X-ray luminosity of $10^{32}$ erg s$^{-1}$ and no X-rays are found for H$_3^+$ with an X-ray enhancement of a factor of 6. The abundances of NH, C and HCO$^+$ are enhanced by a factor of about three. Compared to the spherically symmetric model, C and HCO$^+$ are enhanced on a similar level, while NH is much less enhanced. Destruction of CO$_2$ and water by X-rays has been found in the spherical models of \citet{Staeuber05} and \citet{Staeuber06}, respectively. Since the abundances of H$_2$O and CO$_2$ are already decreased by FUV radiation, this effect is not seen in the two-dimensional model.

The extra enhancement of diatomic hydrides due to X-rays over FUV processing is much smaller in the two-dimensional models compared to the spherical model. As discussed in Section \ref{sec:chemflowwall}, the enhancement in the outflow walls by FUV is much larger than by X-rays. Only the volume averaged abundance NH, which is X-ray sensitive but not much enhanced by FUV radiation shows some dependence on X-ray irradiation (Figure \ref{fig:abus_std1}). However note that NH has high energy levels and thus is more excited in the outflow walls so that the effect of X-rays on the line flux is probably very small. However, a detailed calculation of the line flux should be carried out once collision rates of NH have become available.

The X-ray enhancement/destruction in different models is studied in Table \ref{tab:xrayenh} by the ratio of the volume averaged abundance including protostellar X-ray emission ($L_{\rm X} = 10^{32}$ erg s$^{-1}$) and no X-rays within 4000 AU. The X-ray enhanced species H$_3^+$, NH and HCO$^+$ are similarly enhanced for the 2D models except for models \textit{$\alpha$80}, \textit{$\alpha$100} and \textit{Disk} where the enhancement in smaller. Models \textit{$\alpha$80} and \textit{$\alpha$100} have less mass with only X-rays but no FUV irradiation (Section \ref{sec:physstruc}) and the molecules are photodissociated in the region with X-ray irradiation. In the \textit{Disk} model, however, these molecules are less FUV dissociated due to the higher density and even enhanced in the warm disk atmosphere. This enhanced layer dominates over the X-ray enhanced region and thus the X-ray enhancement is smaller.

X-ray destruction of H$_2$O and CO$_2$ (see \citealt{Staeuber05,Staeuber06}) is remarkable in models which have a larger region with temperature above 100 K but no FUV irradiation (Models \textit{Spherical}, \textit{$\alpha$20}, \textit{$\alpha$40}, \textit{$\gamma$0.1} and \textit{L4e4}). Models with a smaller or similar amount of H$_2$O or CO$_2$ behave similarly to the model \textit{Standard}, discussed before.

\begin{table}[tbh]
\caption{Enhancement or destruction by X-rays for selected species. The ratio of the volume averaged abundance within 4000 AU between a model with $L_{\rm X} = 10^{32}$ erg s$^{-1}$ and no protostellar X-ray emission is shown ($\langle X \rangle_{4000}^{L_{\rm X} = 10^{32}} / \langle X \rangle_{4000}^{L_{\rm X} = 0}$).\label{tab:xrayenh}}
\centering
\begin{tabular}{lll|ccccc}
\hline
  &  Model                &  $M_{4000}$ & H$_3^+$ & NH & CO$_2$ & H$_2$O & HCO$^+$ \\
\hline
  & \textit{Spherical}    &  2.1   &       6.6    &    12      &    0.08          &  0.1            & 4.0           \\
  & \textit{Standard}     &  1     &       5.6    &    1.8     &    1.0           &  1.6            & 3.5           \\ 
  & \textit{Disk}         &  9.1   &       1.8    &    1.1     &    0.7           &  0.7            & 1.1           \\
  & \textit{$\alpha$20}   &  1.8   &       5.9    &    2.8     &   0.01           &  0.1            & 4.1           \\
  & \textit{$\alpha$40}   &  1.5   &       6.0    &    2.2     &   0.03           &  0.4            & 4.4           \\
  & \textit{$\alpha$80}   &  0.64  &       4.2    &    1.2     &    1.5           &  1.3            & 1.7           \\
  & \textit{$\alpha$100}  &  0.36  &       2.6    &    1.0     &    1.9           &  1.0            & 1.0           \\
  & \textit{$\gamma$0}    &  1     &       5.9    &    2.0     &    0.9           &  1.5            & 3.7           \\
  & \textit{$\gamma$0.01} &  1     &       5.5    &    1.9     &    0.5           &  1.5            & 3.7           \\
  & \textit{$\gamma$0.1}  &  1     &       6.6    &    1.9     &   0.05           &  1.0            & 5.2           \\
  & \textit{b1.5}         &  1.3   &       5.3    &    1.6     &   0.03           &  0.4            & 4.0           \\
  & \textit{b2.5}         &  0.76  &       5.4    &    1.6     &    1.4           &  2.2            & 2.6           \\
  & \textit{L2e3}         &  1     &       6.7    &    2.1     &    2.1           &  1.6            & 3.5           \\
  & \textit{L1e4}         &  1     &       5.6    &    1.6     &    1.3           &  1.6            & 3.6           \\
  & \textit{L4e4}         &  1     &       5.3    &    2.0     &   0.05           &  1.3            & 3.3           \\
\hline
\end{tabular}
\end{table}

We conclude that the influence of X-rays on the chemical composition is relatively small in this scenario of FUV irradiated outflow walls. Molecules predicted to be X-ray tracers (\citealt{Staeuber05}) and in particular diatomic hydrides are much more enhanced by FUV radiation compared to X-rays. Exceptions are the water and CO$_2$ destruction by X-rays which is seen in some of the models that are similar to spherical geometry (e.g. small outflow opening angle). Also, HCO$^+$ and H$_3^+$ are considerably enhanced for some models. The differences between the models and the uncertainties of the chemical model are however as large as the extra enhancement by X-rays. For example the total amount of HCO$^+$ (obtained from Tables \ref{tab:avgmods1}, \ref{tab:avgmods2} and \ref{tab:xrayenh}) within 4000 AU is larger for the model \textit{$\alpha$40} without X-rays compared to the model \textit{$\alpha$80} with X-rays. Thus, it is difficult to establish X-ray tracers for these models with very strong FUV irradiation.

In envelopes of low-mass YSOs, however, FUV radiation could be less dominant compared to X-rays. Several effects control the relative importance of FUV and X-rays. The colder photosphere of low-mass YSOs emits less FUV. Bow shocks, internal jet shocks and accretion disk boundary layer can however still produce a considerable FUV field (\citealt{vanKempen09c,vanKempen09b}). On the other hand, most commonly observed low-mass YSOs are closer. Since the X-ray luminosity of low-mass YSOs is of similar strengths compared to high-mass YSOs (e.g. \citealt{Nakajima03}) the X-ray irradiated gas may fill the beam and thus the influence of X-rays on observable molecular lines becomes stronger. A detailed discussion is warranted. Very high resolution interferometric observations of FUV and X-ray tracer with ALMA will also help to disentangle the effects of FUV and X-ray irradiation. In particular the X-ray tracer NH, discussed before, has a line at 946 GHz within ALMA band 10. High-$J$ HCO$^+$ lines (e.g. the $J=10 \rightarrow 9$ line at 892 GHz) which probe the innermost part of the envelope are also within ALMA band 10.

\section{Molecular excitation and line fluxes} \label{sec:molline}

Observable quantities of the models presented here are molecular or atomic line fluxes. In this section, we study different excitation effects of diatomic hydrides, like collisional or radiative pumping. We use a multi-zone escape probability method described in Appendix \ref{sec:molexcit}. Herschel observations are then simulated from the excitation calculation. For simplicity, we restrict the discussion to two prototypical diatomic hydrides, one found to be destroyed (SH) and enhanced (CH$^+$) in the outflow wall. Line fluxes of other species will be presented in an further paper together with observations.

Molecular data for the line frequency and Einstein coefficients are obtained from the CDMS (\citealt{Mueller01}) for CH$^+$ and the JPL database (\citealt{Pickett98}) for SH. The molecular collision rates of many diatomic hydrides are poorly known and extrapolation is thus  necessary. For SH-H$_2$, we adopt the OH-H$_2$ rates by \citet{Offer92}, scaled for the reduced mass. Since collision rates for ortho and para H$_2$ exist, we use a thermal ortho to para ratio\footnote{${\rm oH}_2/{\rm pH}_2 = \min(3,9 \times \exp(-170.6/T_{\rm gas}))$}. For CH$^+$-H$_2$ and CH$^+$-H, we scale the CH$^+$-He rates by \citet{Hammami08} and \citet{Hammami09}. As CH$^+$ is abundant in regions with electron fraction of order $10^{-4}$, we also consider excitation by electron impact (\citealt{Lim99b}). A general discussion on the uncertainty of collision rates is given in \citet{Schoier05}.

Transitions observable by Herschel (HIFI/PACS) are summarized in Tables \ref{tab:moldata_sh} and \ref{tab:moldata_chplus}. They give the transition wavelength or frequency ($\nu_{ul}$, $\lambda_{ul}$), the Einstein-A coefficient $A_{ul}$, the energy of the upper level $E_{ul}$. Transitions of CH$^+$ with $J_{\rm up}=2-6$ can be observed by PACS, while the $J=2\rightarrow 1$ and $J=1\rightarrow 0$ lines at 835.079 GHz and 1669.170 GHz, respectively, are accessible to HIFI. The SH lines are in HIFI bands,  except for the lines at 1383 GHz, which are given for completeness. For SH, only the strongest hyperfine components are used and overlapping components have been combined. The critical density ($n_{\rm crit} = A_{ul} / \sum_l K_{ul}$) gives the necessary density of collision partners for substantial population of the upper level.

\begin{table}[tbh]
\caption{Molecular data of SH. The critical density is given for a temperature of 100 K and a thermal ortho to para ratio of H$_2$.\label{tab:moldata_sh}}
\centering
\begin{tabular}{lrrrr}
\hline
& $\nu_{ul}$ & $A_{ul}$ & $E_{ul}$ & $n_{\rm crit}({\rm H}_2)$ \\
{}[J,F/Parity]           & [GHz] & [s$^{-1}$] & [K] & [cm$^{-3}$] \\
\hline
${}^2\Pi_{1/2}$  $3/2,1- \rightarrow 1/2,1+$ &   866.947  &  1.9(-3)   &       571    &   4.3(6) \\
${}^2\Pi_{1/2}$  $3/2,2+ \rightarrow 1/2,1-$ &   875.267  &  1.5(-3)   &       572    &   4.7(6) \\
${}^2\Pi_{3/2}$  $5/2,3+ \rightarrow 3/2,2-$ &  1382.910  &  9.4(-3)   &        66    &   3.8(7) \\
${}^2\Pi_{3/2}$  $5/2,3- \rightarrow 3/2,2+$ &  1383.241  &  9.4(-3)   &        66    &   4.7(7) \\
${}^2\Pi_{1/2}$  $5/2,2+ \rightarrow 3/2,1-$ &  1447.012  &  1.6(-3)   &       641    &   2.6(7) \\
${}^2\Pi_{1/2}$  $5/2,3- \rightarrow 3/2,2+$ &  1455.100  &  1.6(-3)   &       642    &   3.0(7) \\
${}^2\Pi_{3/2}$  $7/2,4- \rightarrow 5/2,3+$ &  1935.206  &  3.5(-2)   &       159    &   1.2(8) \\
${}^2\Pi_{3/2}$  $7/2,4+ \rightarrow 5/2,3-$ &  1935.847  &  3.5(-2)   &       159    &   1.4(8) \\
\hline
\end{tabular}
\end{table}

\begin{table}[tbh]
\caption{Molecular data of CH$^+$. The critical density is given for a temperature of 500 K.\label{tab:moldata_chplus}}
\centering
\begin{tabular}{lrrrrrr}
\hline
  & $\lambda_{ul}$ & $A_{ul}$ & $E_{ul}$ & $n_{\rm crit}({\rm H}_2)$ & $n_{\rm crit}({\rm H})$ & $n_{\rm crit}(e^-)$ \\
{}[J]            & [$\mu$m] & [s$^{-1}$] & [K] & [cm$^{-3}$]& [cm$^{-3}$]& [cm$^{-3}$]\\ 
\hline
$1  \rightarrow 0$ &    359.0 &  6.4(-3)   &    40     &   5.3(7) & 3.8(7)  & 3.0(4) \\
$2  \rightarrow 1$ &    179.6 &  6.1(-2)   &   120     &   2.5(8) & 1.9(8)  & 1.3(5) \\
$3  \rightarrow 2$ &    119.9 &  2.2(-1)   &   240     &   7.1(8) & 5.2(8)  & 3.8(5) \\ 
$4  \rightarrow 3$ &    90.0  &  5.4(-1)   &   400     &   1.6(9) & 1.2(9)  & 7.8(5) \\
$5  \rightarrow 4$ &    72.1  &  1.1(0)    &   600     &   3.2(9) & 2.3(9)  & 1.7(6) \\
$6  \rightarrow 5$ &    60.2  &  1.9(0)    &   838     &   5.8(9) & 4.2(9)  & 3.0(6) \\
\hline
\end{tabular}
\end{table}

We use wavelength ($\mu$m) for CH$^+$ as most lines are observable with PACS and frequency (GHz) for SH (HIFI). Levels of CH$^+$ are labeled by the rotational quantum number $J$. The molecular structure of SH is similar to hydroxyl (OH)  with a ${}^{2}\Pi$ electronic ground-state, $\Lambda$-type doubling and hyperfine splitting. Levels are thus denoted by ${}^{2}\Pi_\Omega, J,F+/-$, with the total angular momentum $\Omega$, the rotational angular momentum $J$, the nuclear spin $F$ and the parity (+/-). The CH$^+$ and SH molecules show typical properties of diatomic hydrides, with high critical densities ($>10^6$ cm$^{-3}$ for H$_2$ collisions) and large energy separation between subsequent levels. This is due to the large rotation constant combined with a large dipole moment. Line frequencies thus mostly lie above the atmospheric windows. Excitation mechanism other than collisional excitation are likely. For example, pumping by dust continuum radiation or formation in an excited state have been suggested by \citet{Black98} for CH$^+$.

\subsection{Excitation: Level population} \label{sec:molexci}

\begin{figure*}[tbh]
  \centering
  \includegraphics[width=1.0\hsize]{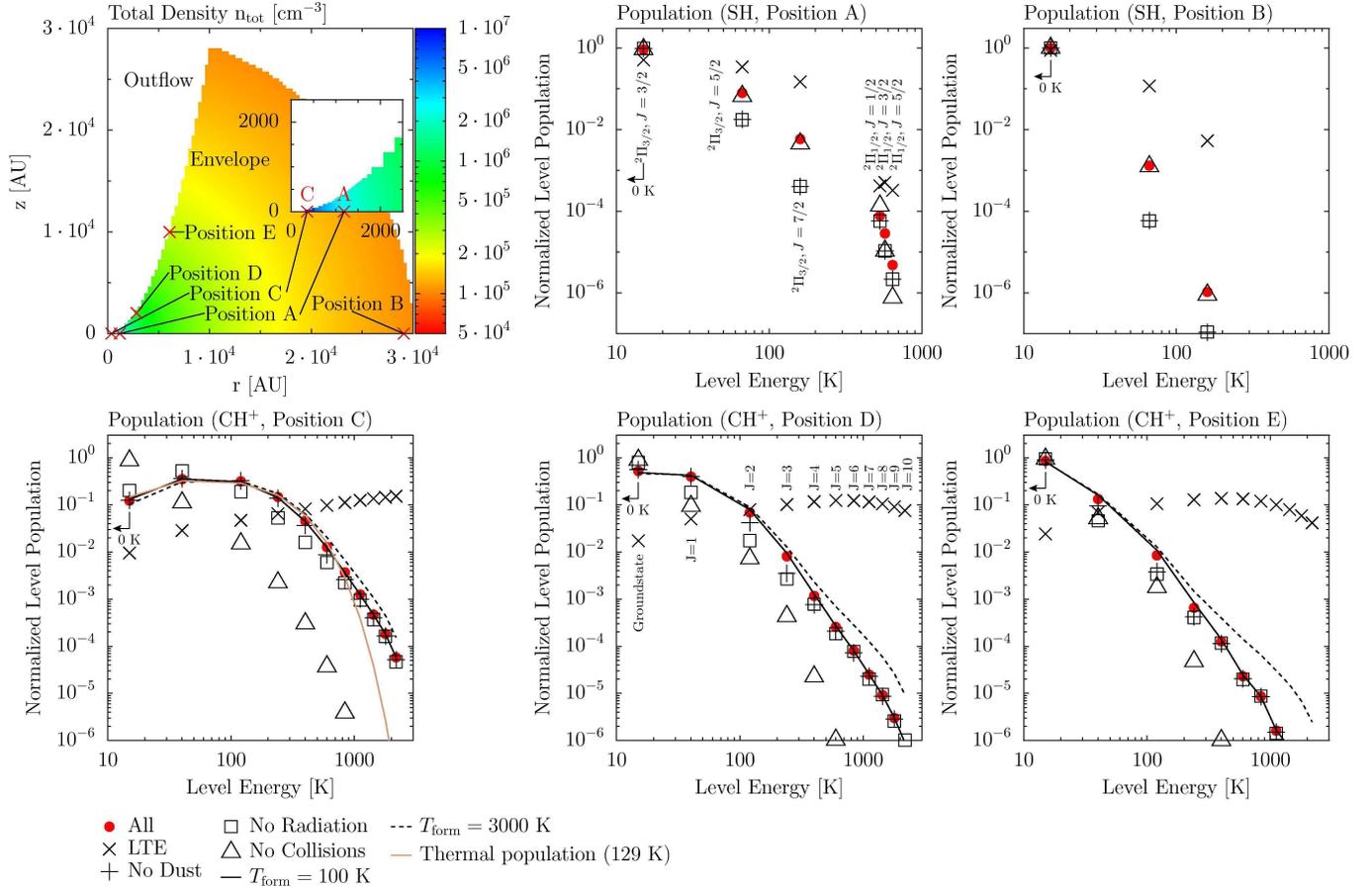}
  \caption{Normalized level population of SH (top panels) and CH$^+$ (bottom panels) at different positions. The positions A-E are given in the density plot in the top/left panel. A zoom-in of the innermost region is given as inset. The level energy of the ground state has been shifted from 0 K to 15 K for the representation in the logarithmic plot.}
  \label{fig:plot_exci}
\end{figure*}

\begin{figure}[tbh]
  \centering
  \includegraphics[width=1.0\hsize]{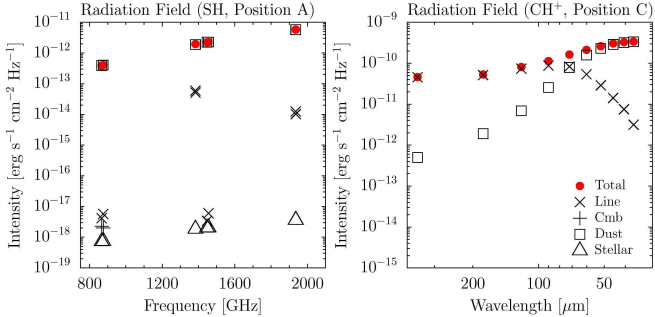}
  \caption{Radiation field of SH and CH$^+$ at positions A and C, respectively. The positions are given in Figure \ref{fig:plot_exci}.}\label{fig:plot_excirad}
\end{figure}

The calculated level populations of SH and CH$^+$ at different positions of model \textit{Standard} are given in Figure \ref{fig:plot_exci}. For SH, the position in the midplane with peak abundance and at the edge of the cloud are presented. For CH$^+$, we show positions with peak abundances along cuts through the midplane and at $z=2000$ AU and $z=10000$ AU (Section \ref{sec:chemstd}). In order to study the importance of different excitation mechanism, we give the normalized level populations depending on the upper level energy for the following models:
\begin{itemize}
\item \textit{All}. Collisional and radiative excitation considered.
\item \textit{No Dust}. Dust radiation switched off by setting the dust opacity to zero. This model allows to study the importance of dust pumping.
\item \textit{No Radiation}. Ambient radiation field switched off ($\langle J_{ij} \rangle=0$). This model together with the previous allows to study the importance of pumping by line emission.
\item \textit{No Collisions}. Density of the collision partners set to zero. This situation corresponds to pumping purely by dust radiation.
\item \textit{LTE}. Level population in the local thermal equilibrium (LTE) with the gas temperature. This model is given for comparison as the assumption of LTE is often made due to missing collision rates.
\item \textit{$T_{\rm form}=100$ K,  $T_{\rm form}=3000$ K (Only CH$^+$)}. Collisional excitation, radiative excitation and excited formation considered with formation temperature $T_{\rm form}$ (Appendix \ref{sec:molexcit}). We adopt a $T_{\rm form}$ of order of the kinetic temperature (3000 K) or much lower (100 K).
\end{itemize}
For SH, the population of the hyperfine quadruplet is combined. We note that the level population can only be translated directly to emissivity per volume in the case of optically thin radiation. Figure \ref{fig:plot_excirad} gives different components (line emission, dust emission, cosmic microwave background and stellar photons) of the ambient radiation field $\langle J_{ij} \rangle$ at positions A (SH) and C (CH$^+$).

The CH$^+$ level population is far out of LTE. It peaks in the $J=1$ state in position C and the ground state in positions D and E. For all positions, the population rapidly decreases to higher $J$ levels and from position C to E for the same level. Collisional excitation (model \textit{No Collisions}) governs the excitation with increasing differences to model \textit{All} for higher $J$. A model without electron excitation but collisions with H and H$_2$ has level populations within 20 \% to the model \textit{All}. Electron excitation is thus unimportant relative to other effects. Models \textit{No Dust} and \textit{No Radiation} differ less relative to model \textit{All} in CH$^+$ compared to SH, thus radiation is less important. Unlike SH, the models \textit{No Dust} and \textit{No Radiation} differ, due to the contribution of line photons to the ambient radiation field. At position C, line photons dominate the radiation field up to the $J=5 \rightarrow 4$ transition (Figure \ref{fig:plot_excirad} right). Optical depths of up to $\tau \approx 1$ (line and dust) for vertical cuts through the outflow wall are reached in the lower $J$ transition. For levels connected by these transitions, we find the largest differences between models \textit{No Dust} and \textit{No Radiation}. We conclude that CH$^+$ is mainly collisionally excited with small contributions from pumping by line trapping and dust radiation.

The importance of collisions for the excitation of CH$^+$ despite the density of the collision partners below the critical density can be understood by the high kinetic temperature. Radiative transitions connect only levels with $\Delta J = 1$ due to selection rules. Rates for radiative pumping are proportional to the lower level population of a pumping transition and thus decrease quickly to higher $J$. Collisional excitation is not restricted to $\Delta J=1$ and can pump over several levels, especially for high kinetic temperatures, when the upward rates are similar or even higher than the downward rates ($C_{lu} g_l = C_{ul} g_u \exp(-\Delta E / k T_{\rm kin})$).

At position C in the model \textit{Standard}, the radiative excitation rate decreases from about $10^{-2}$ ($J=1\rightarrow 2$) to $10^{-5}$ cm$^{-3}$ s$^{-1}$ ($J=9\rightarrow 10$), for $B_{lu} \langle J_{ul} \rangle$ and $B_{ul} \langle J_{ul} \rangle$ of order 0.1 s$^{-1}$ and a CH$^+$ density of 0.4 cm$^{-3}$. The collisional excitation rates from the most populated levels ($J\leq 3$) to a higher level is found to be about $5 \times 10^{-3}$ s$^{-1}$ cm$^{-3}$ into $J=4$ and slightly less for higher $J$, for rate coefficients $K_{ul}$ of order a few times $10^{-10}$ cm$^{3}$ s$^{-1}$ and a density of $10^7$ cm$^{-3}$. Thus, for $J=4$, radiative and collisional rates are similar, but collisions dominate for higher $J$. This is in agreement with the findings in the previous paragraphs.

Excited formation affects the level population relatively little with considerable differences only for higher $J$ levels at lower densities (position E) and a high formation temperature ($T_{\rm form} = 3000$ K). At position C, the total formation rate of CH$^+$ is $6 \times 10^{-3}$ s$^{-1}$ cm$^{-3}$, larger than the collisional excitation, but the effect on the level population is small due to the distribution over different levels following the Boltzmann statistic. Due to this distribution, which supports the thermalization to the formation temperature, the influence of excited formation is mostly seen in higher $J$ levels for $T_{\rm form} = 3000$ K, where collisional excitation competes less with molecules formed in an excited state.

Observation of the $J=2 \rightarrow 1$ up to $J=6 \rightarrow 5$ of CH$^+$ using the Infrared Space Observatory (ISO) have been reported by \citet{Cernicharo97} towards the planetary nebula NGC 7027. They can be explained by a single excitation temperature of $150\pm 20$ K for a kinetic temperature of about 300-500 K and densities of a few times $10^7$ cm$^{-3}$. It is interesting to note that, at position C of our model, the level population up to $J=6$ can be well fitted by a single Boltzmann distribution with a temperature of 129 K. Note that \citet{Cernicharo97} have used N$_2$H$^+$ collision rates and a slightly higher density. Our  calculation thus suggests that the observed line ratio can be explained even with a much higher kinetic temperature as is required for the formation. A more detailed model of their source is however required to draw further conclusions.

The SH level population is out of LTE like CH$^+$. The molecule is however destroyed by FUV radiation and thus abundant in regions with lower temperatures compared to CH$^+$. The level population is thus concentrated in the ground state for both positions A and B. Levels of the ${}^{2}\Pi_{1/2}$ ladder with level energy above 500 K are sparsely populated with relative population below $10^{-4}$. The agreement of the model \textit{No Collisions} with the model \textit{All} in the ${}^{2}\Pi_{3/2}$ states indicates that these level are pumped by dust radiation. The ${}^{2}\Pi_{1/2}$ states on the other hand are connected by lines with lower frequency and thus weaker dust radiation field. These levels are pumped by both collisions and dust radiation.

\subsection{Line fluxes} \label{sec:lineflux}

\begin{figure*}[tbh]
  \centering
  \includegraphics[width=0.75\hsize]{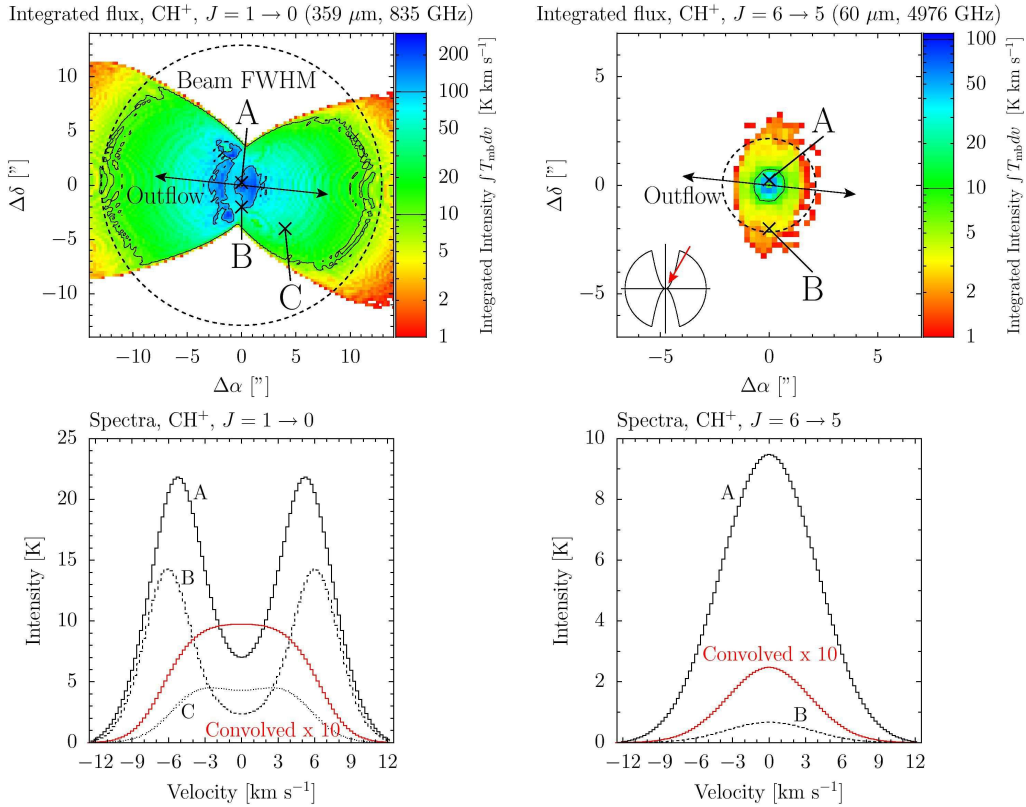}
  \caption{Synthetic velocity integrated maps (top panels) and spectra (bottom panels) of the CH$^+$ $J=1\rightarrow 0$ (left panels) and $J=6 \rightarrow 0$ transitions (right panels) for the model \textit{Standard}. The appropriate Herschel beam (FWHM) is given by a circle. Contour lines give integrated intensities of 10 and 100 K km s$^{-1}$. The continuum is subtracted from the spectra and the red (gray) spectrum is convolved to the Herschel beam and multiplied by 10. The spectra at positions A, B and C are given in a pencil beam.\label{fig:mapspec}}
\end{figure*}

Calculated line fluxes of SH and CH$^+$ for different models are given in Table \ref{tab:mol_flux}. Synthetic frequency integrated maps and spectra at selected positions for the  $J=1 \rightarrow 0$ and $J=6 \rightarrow 5$ lines of the model \textit{Standard} are shown in Figure \ref{fig:mapspec}. Maps and spectra are obtained from the level population discussed in the previous section. Synthetic maps are calculated by integrating the radiative transfer equation. To simulate observations, the maps are convolved with the appropriate Herschel beam, centered on the source and assumed to be a Gaussian. For the raytracing, we assume the same physical parameters as in paper II. These are a distance of 1 kpc, an inclination of $30^\circ$ and a position angle of the outflow direction of $96^\circ$.

\begin{table*}[tbh]
\tablewidth{0pt}
\footnotesize
\caption{Molecular line fluxes of SH and CH$^+$.\label{tab:mol_flux}}
\centering
\begin{tabular}{l|cccccc}
\tableline
 Model            & Line Flux \\
& [K km s$^{-1}$]$^b$ & [K km s$^{-1}$] & [K km s$^{-1}$] & [K km s$^{-1}$] & [K km s$^{-1}$] & [K km s$^{-1}$] \\
\tableline
\textbf{CH$^+$}  & $1\rightarrow 0$ & $2\rightarrow 1$ & $3\rightarrow 2$ & $4\rightarrow 3$ & $5\rightarrow 4$ & $6\rightarrow 5$ \\
Wavelength [$\mu$m]     & 359.0& 179.6 & 119.9 & 90.0 & 72.1 & 60.2 \\ 
Herschel-Beam [''] & 25.8'' & 12.9'' & 8.6'' & 6.5'' & 5.2'' & 4.3'' \\
\tableline
\textit{Standard$^a$}                     &      12.3 &      23.9    &   20.2   &   10.5  &    4.6 &     2.1 \\
\textit{Slab at LTE$^c$}                  &       9.8 &      22.5    &   22.4   &   13.6  &    5.3 &     1.6 \\
\textit{Standard (2 kpc)}                 &       4.3 &       8.9    &    8.0   &    4.2  &    1.9 &     0.9 \\
\textit{Standard (LTE)}                   &      12.0 &     113.1    &  333.0   &  612.3  &  871.9 &  1064.9 \\
\textit{Standard ($T_{\rm form}=100$ K)}  &      14.9 &      26.4    &   21.8   &   11.0  &    4.7 &     2.1 \\
\textit{Standard ($T_{\rm form}=3000$ K)} &      15.8 &      30.8    &   29.4   &   18.1  &    9.7 &     5.4 \\
\textit{Standard (0$^\circ$)}             &      14.8 &      29.5    &   26.1   &   14.1  &    6.3 &     3.0 \\
\textit{Standard (90$^\circ$)}            &      16.1 &      23.6    &   17.3   &    8.0  &    3.2 &     1.3 \\
\textit{Standard (25.8'')}                &      12.3 &       7.5    &    2.9   &    0.9  &    0.3 &     0.1 \\
\textit{Standard (Noturb)}                &       9.4 &      19.0    &   19.3   &   11.2  &    5.0 &     2.2 \\
\textit{Disk}                             &       8.2 &      11.7    &    7.3   &    2.6  &    0.8 &     0.3 \\
\textit{$\alpha$20}                       &       4.9 &	     10.1    &   10.7   &    7.2  &    3.5 &     1.6 \\
\textit{$\alpha$100}                      &      28.8 &	     40.7    &   29.8   &   14.6  &    6.7 &     3.3 \\
\tableline
\textbf{SH$^e$}  & ${}^2\Pi_{1/2 (3/2 \rightarrow 1/2)}$ & ${}^2\Pi_{3/2 (5/2 \rightarrow 3/2)}$ & ${}^2\Pi_{1/2 (5/2 \rightarrow 3/2)}$ & ${}^2\Pi_{1/2 (5/2 \rightarrow 3/2)}$ & ${}^2\Pi_{3/2 (7/2 \rightarrow 5/2)}$ & ${}^2\Pi_{3/2 (7/2 \rightarrow 5/2)}$ \\
Frequency [GHz] &  866.947 &    1382.910  &    1447.012  &  1455.100  &  1935.206  &   1935.847 \\
Herschel-Beam [''] & 24.9'' & 15.6'' & 14.9'' & 14.8'' & 11.1'' & 11.1'' \\
\tableline
\textit{Standard$^a$}                     &      3.8(-5) & -2.1(-3)$^d$ &      4.6(-5) &    5.0(-5)   &    5.8(-4) &    4.4(-4) \\
\textit{Standard (2 kpc)}                 &      3.2(-6) &     -9.7(-4) &      2.1(-6) &    2.4(-6)   &    8.7(-5) &    6.1(-5) \\
\textit{Standard (LTE)}                   &      8.2(-5) &      3.2(-1) &      8.8(-4) &    8.5(-4)   &    9.1(-2) &    9.0(-2) \\
\textit{Standard (0$^\circ$)}             &      8.4(-5) &      5.4(-3) &      1.2(-4) &    1.3(-4)   &    9.2(-4) &    7.3(-4) \\
\textit{Standard (90$^\circ$)}            &      9.9(-6) &     -6.1(-2) &      5.7(-6) &    6.6(-6)   &   -7.9(-4) &   -7.6(-4) \\
\textit{Disk}                             &      1.8(-5) &     -9.9(-4) &      2.4(-5) &    2.6(-5)   &    2.8(-3) &    2.3(-3) \\
\textit{$\alpha$20}                       &      1.6(-4) &     -1.5(-1) &      1.8(-4) &    2.0(-4)   &    3.8(-2) &    3.0(-2) \\
\textit{$\alpha$100}                      &      8.2(-5) &      1.7(-3) &      9.8(-5) &    1.1(-4)   &    2.8(-4) &    2.1(-4) \\
\tableline
\end{tabular}
\begin{flushleft}
\footnotesize{
Note: $a(b)=a \times 10^b$.\\
$^a$If no other indication is given, the inclination is 30$^\circ$ and the distance 1 kpc (see Appendix \ref{sec:molexcit})\\
$^b$The conversion factor from K km s$^{-1}$ to W cm$^{-2}$ is $2.7 \times 10^{-14} / (\lambda [\mu{\rm m}])^{3}$.\\
$^c$Fit to model \textit{Standard} for a CH$^+$ column density of $7 \times 10^{12}$ cm$^{-2}$ and $T_{\rm ex}=148$ K.\\
$^d$Negative fluxes mean line in absorption.\\
$^e$See Table \ref{tab:moldata_sh} for the exact label of the transition including fine structure.}
\end{flushleft}
\end{table*}

The predicted CH$^+$ fluxes are remarkably strong, even for higher $J$ lines. The integrated intensities correspond to fluxes of order a few times $10^{-19}$ W cm$^{-2}$ and are easily detectable with PACS or HIFI. Note that they are similar to those reported towards NGC 7027. Strong lines for higher $J$ transitions despite the low level population found in the previous section are due to the smaller beam at shorter wavelengths and the emissivity (in K km s$^{-1}$) per excited molecule $\propto A_{ul}/\nu_{ul}^2$, increasing by a factor of 10 from the $J=1 \rightarrow 0$ to the $J=6 \rightarrow 5$ transition. Using a Boltzmann diagram, corrected for line absorption, the model \textit{Standard} fits well for a column density of $7 \times 10^{12}$ cm$^{-2}$ and an excitation temperature of about 148 K (\textit{Slab at LTE}).

Different radiative transfer models of CH$^+$ derived from the chemical model \textit{Standard} have similar fluxes in the $J=1 \rightarrow 0$ line, with deviation smaller than about 30 \%. An exception is the model \textit{Standard (2 kpc)} which assumes a distance of 2 kpc instead of 1 kpc to the source, as suggested by \citet{Schneider06}. For higher $J$ transitions, fluxes however differ significantly between the models. Assuming the level population in the LTE (\textit{Standard (LTE)}) yields fluxes more than two orders of magnitude larger compared to the model \textit{Standard}. Excited formation increases the line fluxes of all lines by less than 20 \% for $T_{\rm form}=100$ K. Assuming $T_{\rm form}=3000$ K, a larger increase is found for higher $J$ lines and amounts to a factor of 3 for the $J=6\rightarrow 5$ line. Inclination affects the line flux by the combination of line and dust opacity, together with the beam containing different parts of the model. For example, the dust opacity reaches about 1.2 for the $J=6\rightarrow 5$ line, while it is below 0.07 for the $J=1\rightarrow 0$ line. Thus, the flux of the $J=6 \rightarrow 5$ line in the model \textit{Standard (90$^\circ$)} (edge on) is about a factor of two smaller compared to the model \textit{Standard (0$^\circ$)} with a line of sight parallel to the outflow. Compared to the model \textit{Standard}, the inclination changes the flux mostly by less than 40 \%. We note that for other sources with larger total column density, the dust attenuation may be important even for longer wavelengths. Increasing the distance to 2 kpc finally, decreases the fluxes by a factor of about 2.5.

The frequency integrated maps and spectra of model \textit{Standard} given in Figure \ref{fig:mapspec} show that the concentration of the emission to the innermost region is more pronounced for the $J=6\rightarrow 5$ line compared to the $J=1\rightarrow 0$. This is also reflected in a model (\textit{Standard (25.9'')}), convolved to a 25.8'' beam. The flux ratios of that model to model \textit{Standard} decreases almost as fast as the beam dilution factor, indicating that most of the emission is within the beam. The profile of the $J=1\rightarrow 0$ line shows considerable self-absorption in the region where the line of sight follows the outflow wall due to the opening angle being similar to the inclination (positions A and B). The optical depth at the line center at these positions are 7 (A) and 19 (B). Further out, the optical depth is about 3 (C), which is about twice the value found in the Section \ref{sec:molexci} for a ray crossing the outflow wall once. The optical depth is even small for distances larger than position C. The $J=6\rightarrow 5$ line has much smaller line optical depths below 0.2 (position A) compared to the $J=1\rightarrow 0$ line. Beam convolved spectra of both lines however, do not show a self-absorption dip, suggesting that most of the emission comes from regions with moderate optical depth.

The considerable line optical depths for lines of sight along the outflow walls raises the question, how much the modeled line fluxes depend on the assumed intrinsic line shape and velocity structure. HIFI can observe the $J=1\rightarrow 0$ and $J=2\rightarrow 1$ lines of CH$^+$ spectrally resolved but with a beam corresponding to more than 10000 AU at the assumed distance of 1 kpc. This is insufficient to directly probe the velocity structure along the outflow wall. It can however be used to constrain the average line width along the outflow wall. Spatially resolved interferometric observations reveal broader lines in the outflow wall without much velocity shift to the systemic velocity (\citealt{Bruderer09c}). This has been implemented in model \textit{Standard} (Appendix \ref{sec:molexcit}). A model without increased microturbulence in the outflow walls (\textit{Standard (Noturb)}) has been calculated and shows total line fluxes within 20 \% to model \textit{Standard}. We conclude that the while the modeled line shape may depend strongly on the assumed intrinsic line and velocity structure, the line fluxes are pretty independent.

Different chemical models (\textit{$\alpha$20}, \textit{$\alpha$100} and \textit{Disk}) affect the line fluxes of CH$^+$ by a combination of the chemical abundance, the temperature and density structure and also the beam containing different regions of the model. For example the total amount of CH$^+$ in the models \textit{$\alpha$20}, \textit{$\alpha$100} and \textit{Disk} within 4000 AU relative to the model \textit{Standard} are 0.64, 0.36 and 1.04, respectively (Tables \ref{tab:avgmods1} and \ref{tab:avgmods2}). The line fluxes relative to model \textit{Standard} are thus similar to the abundance ratios for models \textit{$\alpha$20} and \textit{Disk} in low-$J$ lines. For \textit{$\alpha$100}, the line is however much stronger due to the larger mass with high temperature (Table \ref{tab:propafgl}). For higher $J$ lines, the situation is different. The lines of the model \textit{Disk} get weaker with $J$ relative to model \textit{Standard}, while those of model \textit{$\alpha$20} get stronger. This is due to the fact, that the abundance of CH$^+$ decreases in model \textit{Disk} in regions with high density, while in the model \textit{$\alpha$20} more material with high density is heated by FUV. We conclude that the low-$J$ lines of CH$^+$ are better tracers for the chemical abundance than higher $J$ lines, as the higher $J$ lines depend more on the temperature structure. The difference between the models is of order of about 3, similar to the uncertainty of the chemical model. Details of the geometry can thus not be obtained solely by observations of diatomic hydrides.

The SH lines are very weak, with most line fluxes clearly below the detection limit. This is remarkable since the volume averaged abundance of SH (Table \ref{tab:avgstd1d2d}) is only about two orders of magnitude smaller than for CH$^+$. The weak fluxes can be explained by the poor excitation in regions with high abundance of SH, due to the large energy separation of the levels (Section \ref{sec:molexci}). Lines of the ${}^{2}\Pi_{1/2}$ ladder with upper level energies of several 100 K have line fluxes below $10^{-3}$ K km s$^{-1}$. Lines between ${}^{2}\Pi_{3/2}$ states can reach line fluxes of a few times $10^{-2}$ K km s$^{-1}$ owing to their level energies similar to the dust or gas temperature in regions where SH is abundant. These lines can also be seen in absorption if a sufficient column density of cold dust lies between the SH emitting region and the observer. Unfortunately, this feature is most prominent in the ${}^{2}\Pi_{3/2(5/2\rightarrow3/2)}$ line at 1383 GHz which cannot be observed by PACS or HIFI.

The abundance of SH within 4000 AU of the models \textit{$\alpha$20}, \textit{$\alpha$100} and \textit{Disk} relative to model \textit{Standard} are 180, 0.04 and 23 (Tables \ref{tab:avgmods1} and \ref{tab:avgmods2}). This is only reflected in lines of the ${}^{2}\Pi_{3/2}$ levels which are excited in larger parts of the envelope compared to the ${}^{2}\Pi_{1/2}$ levels. Because of the larger amount of warm material, lines obtained from model \textit{$\alpha$100} can be stronger than from the model \textit{Standard} despite the lower abundance. We conclude that the excitation in the warm region of the FUV outflow walls is essential for observable lines of diatomic hydrides. Thus, observations or non-detections do not necessarily reflect column densities, due to high critical densities combined with large energy separations between subsequent energy levels. 

\section{Observing diatomic hydrides with Herschel and ALMA} \label{sec:herschobs}

The previous sections have shown that light diatomic hydrides (CH, NH, SH, OH) and their ions belong to a particular class of molecules that are expected to probe the outflow walls of YSO envelopes. This is due to three factors relating to formation and excitation of the diatomic hydrides: i.) Owing to considerable activation barriers for the formation of many of the diatomic hydrides, a high temperature of several 100 K is needed for their formation. ii.) The chemical network leading to the formation is initiated by photoionization and large FUV fields are required for the formation of diatomic hydrides. iii.) Due to high critical densities and a large energy separation between the molecular levels, a high density and temperature is needed for excitation of the diatomic hydrides.

Protostellar FUV radiation can escape through a low density cavity, etched out by the outflow, and irradiate the outflow walls separating the outflow and envelope. Our work predicts the abundance and excitation of diatomic hydrides in this thin outflow walls to govern the total line flux and result in fluxes that can be easily detected. Diatomic hydrides thus are predicted to be valuable tracers for the chemistry and physics of the outflow walls, which are in many ways similar to the upper atmosphere of protoplanetary disks.

The Herschel Space Observatory provides a unique opportunity to observe diatomic hydrides, as their lines are above atmospheric windows. Beyond the relatively large sample of hydride lines in the HIFI spectral range, which is targeted by the WISH Herschel guaranteed time key program in this source, PACS observations of higher $J$ lines may yield important constraints on the hottest regions. A particular example are high-$J$ lines of CH$^+$ which are predicted to be detectable by PACS. For this particular source, the relatively large beam of Herschel does not allow to spatially resolve the outflow walls and our model do also not indicate that the predicted line fluxes can be used easily as tracer for the exact shape of the cavity. In other sources with a good balance between distance and affected outflow area (e.g Serpens-FIRS 1), the good spatial resolution of the PACS camera may allow to spatially resolve different transitions of CH$^+$.

The upcoming Atacama Large Millimeter and submillimeter Array (ALMA) can observe diatomic hydrides at very high angular resolution using band 10 receivers (787-950 GHz). For example the $J=1\rightarrow 0$ line of the ${}^{13}{ \rm CH}^+$ isotope is shifted sufficiently to be inside an atmospheric window (\citealt{Falgarone05}). Also OH$^+$ at 909 GHz (\citealt{Wyrowski10}) and SH$^+$ at 893 GHz (\citealt{Menten10}) have been detected from ground. Furthermore, lines of NH (946 GHz) and SH (866 and 875 GHz) are observable with ALMA band 10 receivers. Herschel observations can thus be complemented in the near future with high-resolution interferometric data to constrain parameters like the geometry in the innermost region.

\section{Limitations of the model} \label{sec:limit}

The scenario of directly irradiated outflow walls by radiation of the protostar is not the only possible to alter the chemical composition in the outflow walls. For example mixing between the warm and ionized outflow with the envelope has been suggested by \citet{Taylor95} (see also \citealt{Raga02,Raga07}).  The observed abundance of CO$^+$, however, cannot be explained by mixing solely (paper II). Major dissociative shocks may occur at the boundary layer between outflow and inflow and produce FUV radiation that irradiates the quiescent envelope. To replace the FUV field of direct protostellar radiation with shocks, a rather high shock velocity is required. For example, in the model \textit{Standard} at $z=2000$ AU, $v_s \approx 75$ km s$^{-1}$ is required (\citealt{Neufeld89}). Such conditions are clearly met for the bow shock or internal jet working surfaces. Along the outflow walls, however, the velocity perpendicular to the quiescent envelope is likely smaller. In the simple model of the Couette flow solution (e.g. \citealt{Lizano95}) neglecting turbulence, the mean velocity in the entrainment/mixing is along the outflow cavity and depends linearly on the distance to the cavity edge. The width of this layer and the substructure of the velocity including turbulences are however not easily constrained. Observationally, no clear signs of shocks in the outflow walls have been found in interferometric observations of AFGL 2591 by \citet{Bruderer09c}. We conclude that further study of the combined effects of mixing and shocks is warranted, especially in the context of low-mass star formation with less or cooler protostellar FUV radiation.

Several simplifications have been made to make the calculation feasible. As discussed above, the physical structure does not account for shocks and mixing in the outflow layer or a clumpy surface of the outflow wall. The chemical models used here do not consider non-thermal desorption (e.g. photodesorption). The photodesorption yields are not well known for many species. However, the change in abundance at the 100 K evaporation temperature of H$_2$O, CO$_2$ and H$_2$S is small in the presence of a strong FUV field such as in the outflow wall. Thus the photodesorption would not affect the results of the outflow wall enhanced species considerably. The drop of the water abundance in the region with FUV irradiation but temperatures below 100 K would be mitigated by photodesorption. FUV dissociation and ionization processes are treated here in the rate approach which does not allow to specify the protostellar radiation field explicitly. The surface temperature of the protostar is however not well constrained by observations and would affect the chemistry if the temperature were not hot enough to dissociate CO (\citealt{Spaans95}). The models do not account for vibrationally excited H$_2$ (e.g. \citealt{Sternberg95}) which may help to overcome the activation barrier and further increase the abundance of e.g. CH$^+$ or OH. Grain surface reactions, except for H$_2$ formation and charge exchange with grains, are not considered in the models. For species with longer chemical timescales (chemical clocks, e.g. many sulfur bearing species, \citealt{Wakelam04a}) the effects of infall must be taken into account. 

\section{Conclusions and outlook} \label{sec:concl}

We have used a detailed chemical model of the high-mass star-forming region AFGL 2591 to study the effect of different shapes of a concave low-density cavity on the chemistry. The cavity, etched out by the outflow, allows FUV radiation to escape the innermost region and irradiate the walls between outflow and envelope. We study especially those molecules that are newly observable with the HIFI/PACS instrument onboard Herschel and in particular light diatomic hydrides (CH, NH, SH, OH). The model used in this work is based on the two-dimensional axisymmetric chemical model introduced in \citet{Bruderer09b}, extended by a self-consistent calculation of the dust temperature and an escape probability method to calculate the molecular excitation. The model considers protostellar FUV and X-ray irradiation and makes use of the grid of chemical models presented in \citet{Bruderer09a}. The main conclusions of the work are:
\begin{enumerate}
\item Protostellar FUV radiation can enhance the abundance of light diatomic hydrides and their ions by photoionization/dissociation processes and heating in the outflow-walls considerably. The overall amount of CH$^+$, OH$^+$ and NH$^+$ in the envelope is orders of magnitude larger than in spherically symmetric geometry (Section \ref{sec:chemabu}). Similarly enhanced are C$^+$ and H$_2$O$^+$. These species thus appear to be clear tracers of extended FUV irradiation and thus the geometric structure of the inner regions of YSOs.
\item Depending whether the vertical height of the envelope from the midplane to the outflow (geometrically thin or thick), the size and mass of the infrared heated hot-core region with temperature above 100 K, but no FUV irradiation, can change substantially (Section \ref{sec:physstruc}). Geometrically thin inner regions allows FUV radiation to penetrate to the midplane and destroy the bulk of important species such as CO$_2$ and H$_2$O. The shape of the cavity is thus also important for species which are destroyed by FUV radiation (Section \ref{sec:phyxrayfuv} and \ref{sec:compmods}).
\item Provided that FUV radiation can escape the innermost region, the effects of different geometries (opening angle of the outflow, cavity shape, density in the cavity) on the total amount of species enhanced in the outflow-wall (CH, CH$^+$, OH$^+$, NH$^+$, C$^+$ and H$_2$O$^+$) are relatively small. Deviations are typically less than a factor of three (Section \ref{sec:compmods}).
\item In outflow walls, the influence of protostellar X-ray irradiation on the chemistry and in particular on the diatomic hydrides is relatively small with respect to the much larger enhancement by FUV radiation (Section \ref{sec:xdrivchem}).
\item The large separation between the energy levels of diatomic hydrides requires a hot and dense gas for excitation. Thus, the line fluxes of CH$^+$, predicted to be enhanced in the hot outflow wall, reach fluxes of several K km s$^{-1}$ even in the $J=6\rightarrow 5$ line. On the other hand, SH, is predicted to be destroyed in the outflow wall, but to be abundant in the envelope. Thus, despite the volume averaged abundance of SH being only two orders of magnitude lower than for CH$^+$, it is not sufficiently excited in the cool envelope to be detectable (Section \ref{sec:molexci} and \ref{sec:lineflux}). 
\item Excited formation of CH$^+$ does not considerably affect observable line fluxes. Exceptions are high-$J$ lines, if a large formation temperature is assumed and the ambient density is low (Section \ref{sec:molexci} and \ref{sec:lineflux}).
\end{enumerate}

This suggests that diatomic hydrides are important tracers of warm and FUV irradiated material. Diatomic hydrides are chemically closely related to important species like water. Studying diatomic hydrides will provide important insight in the chemistry of YSO envelopes. Not only the chemistry of the diatomic hydrides is interesting, as they may ultimately also be used a tracers for physical conditions (like the FUV radiation). Herschel will for the first time observe diatomic hydrides with a good sensitivity and angular resolution and making these tracers available. ALMA and in particular band 10 will allow to continue the study of diatomic hydrides at very high angular resolution. Since diatomic hydrides have not been studied thoroughly in the past, much molecular data and in particular collisional excitation rates are missing. More studies on these processes will be necessary in order to analyse Herschel observations.

In this series of papers (\citealt{Bruderer09a,Bruderer09b} and this work), we have introduced a fast method for chemical modeling of envelopes of young stellar objects and studying the molecular excitation. The numerical speed up allows to construct more detailed models including multidimensional geometries. The method will also facilitate the study of larger samples of observations as they will soon be available by new facilities with high sensitivity and larger spectral coverage as for example the upcoming Atacama Large Millimeter/submillimeter Array. Ultimately, extensions of the methods presented here may also be useful to study the chemical/physical evolution of protoplanetary disks.

\begin{acknowledgments}
We thank Ewine van Dishoeck, Susanne Wampfler, Cecilia Ceccarelli, Michiel Hogerheijde and our referee, Tim van Kempen, for useful discussions. The publicly available RATRAN code (\citealt{Hogherheijde00}) has simplified the development of the escape probability code considerably. Submillimeter astronomy at ETH Zurich is supported by the Swiss National Science Foundation grant 200020-121676 (SB, AOB and PS). SDD is supported by the Research Corporation and the NASA grant NNX08AH28G.
\end{acknowledgments}

\clearpage

\begin{appendix}

\section{Dust Radiative Transfer} \label{sec:dustrt}

The dust temperature in steady state condition is obtained by equating the total dust emission $\Gamma_{\rm emit}$ (erg s$^{-1}$ g$^{-1}$) with the total absorption $\Gamma_{\rm abs}$ (erg s$^{-1}$ g$^{-1}$). Dust may absorb either stellar photons or ambient emission by dust. Assuming a single dust temperature $T_{\rm Dust}$, we can write the energy balance at one position of the envelope as
\begin{equation}
\Gamma_{\rm abs} = \int \int \kappa_\lambda I_\lambda(\Omega) d\lambda d\Omega = 4\pi \int \kappa_\lambda B_\lambda(T_{\rm dust}) d\lambda = \Gamma_{\rm emit} \ ,
\end{equation}
with $\kappa_\lambda$ (cm$^2$ g$^{-1}$) the dust absorption opacity per mass, $I_\lambda(\Omega)$ (erg s$^{-1}$ cm$^{-2}$ $\mu$m$^{-1}$ sr$^{-1}$) the intensity depending on the direction and wavelength and $B_\lambda(T_{\rm Dust})$ the Planck function. To solve this coupled problem, we implement the Monte Carlo approach presented by \citet{Bjorkman01}. The results of the new code have been verified with the benchmark tests suggested by \citet{Ivezic97b} and \citet{Stamatellos03} and also by comparing the results with the DUSTY code (\citealt{Ivezic97,Nenkova99}).

\begin{figure}[tbh]
  \centering
  \includegraphics[width=1.0\hsize]{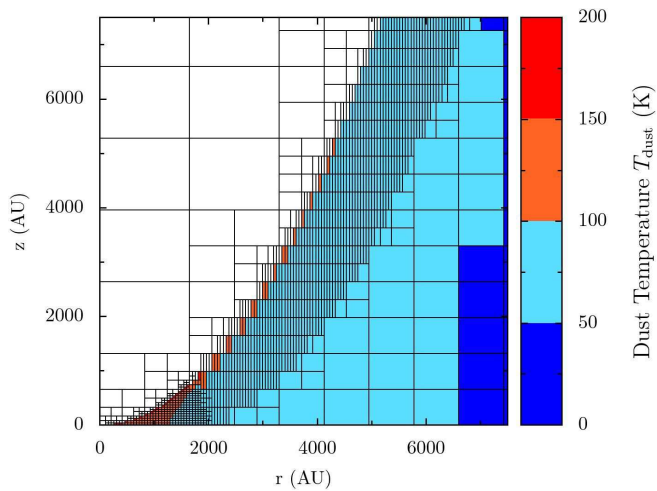}
  \caption{Inner region of the calcalultion grid and the dust temperature of an axisymmetric model of AFGL 2591. \label{fig:dusttempandgrid}}
\end{figure}

For the models of AFGL 2591, we adopt the dust opacities by \citet{Ossenkopf94} (column 5). This dust opacities have been found to explain the observation of AFGL 2591 well and yield a ``standard'' dust to molecule mass ratio of about 100 (\citealt{vdTak99}). For the calculation of the FUV intensity, the dust opacities are extended to shorter wavelengths ($\lambda < 1 \mu$m) with the absorption and scattering properties by \citet{Draine03b}, available on his web-page (similar to paper II). An example of calculated dust temperature is given in Figure \ref{fig:dusttempandgrid} for the model used in paper II. The figure also shows the calculation domain which consists of 5 nested grids.

The dust radiative transfer code also calculates the local mean intensity of scattered and attenuated stellar photons with a Monte Carlo method similar to \citet{vanZadelhoff03}. The analytical expression for the photodissociation and ionization rates (e.g. \citealt{vanDishoeck88}) implemented in paper I require the calculation of the unattenuated FUV flux $G_0$ [ISRF] and the visual extinction $A_V$. We estimate these quantities by the mean intensities at both edges of the FUV band (6 and 13.6 eV) and at 1 micron. We have tested the approach by comparing photoionization and dissociation rates obtained from the fit and calculated by the FUV cross sections of \citet{vanDishoeck06c}\footnote{www.strw.leidenuniv.nl/\~{}ewine/photo}. We find agreement within a factor of two in the chemically relevant range of attenuation ($A_V < 10$). This is reasonable provided that we do not implement the same dust properties and spectral shape of the emitting spectra (see Section \ref{sec:limit} for a discussion of the spectral shape). The FUV calculation is also tested against the results in paper II. We find good agreement, for example the mass and volume of material irradiated with an attenuated flux larger than 1 ISRF agree to within 30 \%.

\section{Molecular Excitation Analysis} \label{sec:molexcit}

For the calculation of the level population of SH and CH$^+$, a multi-zone escape probability method is used. This appendix briefly discusses the method. Further details and the code will be published elsewhere.

The excitation of a molecule or atom (i.e. the population of different energy levels) is controlled by collisional and radiative processes. The species can be excited or deexcited by collisions with H$_2$, H and electrons or by radiation either from the dust continuum or molecular emission from the same or other molecules. The population $n_i$ (cm$^{-3}$) of level $i$ at one position is described by the rate equations (e.g. \citealt{vdTak07})
\begin{equation} \label{eq:rateeqs}
\frac{{\rm d}n_i}{{\rm d}t} = \sum_{j \neq i}^N n_j P_{ji} - n_i \sum_{j \neq i}^N P_{ij} + F_i - n_i D_i \equiv 0 \ ,
\end{equation}
where $N$ is the number of levels considered, $F_i$ (cm$^{-3}$ s$^{-1}$) and $D_i$ (s$^{-1}$) are the chemical formation and destruction rates and $P_{ij}$ (s$^{-1}$) is given by
\begin{equation}
P_{ij} = \left\{
\begin{array}{ll}
A_{ij}+B_{ij} \langle J_{ij} \rangle + C_{ij} & (E_i > E_j) \\
B_{ij} \langle J_{ij} \rangle + C_{ij} & (E_i < E_j) \ . \\
\end{array}
\right. 
\end{equation}
Here, $A_{ij}$ (s$^{-1}$) and $B_{ij}$ (erg$^{-1}$ cm$^2$ Hz) are the Einstein coefficients for spontaneous and induced emission, respectively. The collisional excitation and deexcitation rates $C_{ij}$ (s$^{-1}$) are obtained from the sum of $n_{\rm col} K_{ij}(T_{\rm kin})$ for different collision partners, with the density of the collision partners $n_{\rm col}$ and the collision rate coefficients $K_{ij}(T_{\rm kin})$ (cm$^3$ s$^{-1}$)  at a kinetic temperature $T_{\rm kin}$. The level population is assumed to be constant with time and thus ${\rm d}n_i/{\rm d}t=0$ in Eq. \ref{eq:rateeqs}.

The level distribution at formation may govern the level population if a molecule is destroyed in collisions rather than excited, like CH$^+$ in collisions with H, H$_2$ or electrons. We take the effects of excited formation and destruction into account by the terms $F_i$ and $n_i D_i$ in Eq. \ref{eq:rateeqs}. Due to the short chemical time-scales, we assume the abundance to be constant. At formation, we assume the level population to follow a Boltzmann distribution with temperature $T_{\rm form}$ (\citealt{Staeuber09,vdTak07}). The destruction is assumed to be independent of levels. Hence,
\begin{eqnarray}
F &=& \sum_i F_i= \sum_i F g_i e^{-E_i/kT_{\rm form}}/Q(T_{\rm form}) \nonumber   \\
  &=& \sum_i n_i D_i = D \sum_i n_i = D n({\rm CH}^+) \ , 
\end{eqnarray}
with the partition function $Q(T)$, the statistical weight $g_i$ of a level with energy $E_i$. For the calculation of the destruction rate $D$, we take reactions with H, H$_2$ and electrons into account. Thus $D=k_1 n({\rm H})+k_2 n({\rm H}_2)+k_3 n(e^-)$, with the reaction rate coefficients $k_1$, $k_2$ and $k_3$. In this approach, the formation temperature $T_{\rm form}$ remains a free parameter.

Since the level populations at different positions of a model are coupled by the radiation field, the solution of a radiative transfer problems can be very time-consuming, especially in multi-dimensional geometries or for molecules with high optical depth like water. We thus use the escape probability approach including continuum emission and absorption presented by \citet{Takahashi1983} and approximate the ambient radiation by  
\begin{eqnarray} \label{eq:radescap}
\langle J_{ij} \rangle &\approx& (1 - \epsilon_{ij}) B(T_{{\rm ex},ij},\nu_{ij}) + (\epsilon_{ij} - \eta_{ij}) B(T_{\rm dust},\nu_{ij}) \nonumber \\
&&+\eta_{ij} B(T_{\rm CMB}, \nu_{ij})  \ ,
\end{eqnarray}
with the line frequency $\nu_{ij}$ and the Planck functions $B(T,\nu)$ for the excitation temperature $T_{\rm ex}$, the temperature of the dust $T_{\rm dust}$ and the cosmic microwave background $T_{\rm CMB}$. The escape probabilities $\epsilon_{ij}$ and $\eta_{ij}$ give the probability for a photon to escape line absorption ($\epsilon_{ij}$) or to escape both line and dust absorption ($\eta_{ij}$). They are calculated using Eq. 3.11 - 3.13 given in \citet{Takahashi1983}. The necessary optical depth of dust and lines is obtained from the summation along different rays for different frequencies.

This approach couples different cells and the problem is solved iteratively, similar to one-zone escape probability codes (e.g. RADEX, \citealt{vdTak07}). To improve convergence, we use an ALI-like acceleration mechanism (e.g. \citealt{Rybicki91} or \citealt{Hogherheijde00}). To test the code, we have run several benchmark problems, e.g. \citet{vdTak05b}, with results typically within 30 \% to the exact solution. A similar method has also been successfully used by \citet{Poelman05}.

The velocity field and the intrinsic line width enter the excitation calculation by the ambient molecular radiation field. These parameters are however only important for optically thick lines and the line shape. Optically thin lines are only little affected. The predicted velocity field by \citet{Doty06} (after \citealt{McLaughlin97}) for young chemical ages of less than a few times $10^4$ yrs is approximately static outside a few 1000 AU and we thus assume a static situation. The line width mainly due to microturbulence is assumed to be 1.6 km s$^{-1}$ (Doppler parameter, e.g. \citealt{Staeuber07}) except in the outflow wall, where a larger line width of 4.2 km s$^{-1}$ is assumed as indicated by interferometric observations (\citealt{Bruderer09b}).

\end{appendix}

\bibliographystyle{apj}

\end{document}